\documentclass[11pt,a4paper]{article}

\usepackage[english]{babel}
\usepackage{amsmath}
\usepackage{amssymb}
\usepackage{amsthm,amsfonts}
\usepackage{dsfont}
\usepackage{tensor}
\usepackage[dvipsnames]{xcolor}
\usepackage{slashed}
\usepackage{stmaryrd}
\usepackage{enumerate}
\usepackage[normalem]{ulem}
\usepackage{tikz}
\usepackage[mathscr]{euscript}
\usepackage{mathrsfs}
\usepackage{hyperref}
\usepackage{mathtools}
\usepackage{graphicx}
\usepackage{tcolorbox}
\usepackage{fontawesome}
\usepackage{cite}
\usepackage{pifont}

\usepackage[left=3cm,right=3cm,top=3cm,bottom=3cm]{geometry}

\usepackage{bbold} 
\newcommand\fakeslant[1]{%
  \pdfliteral{1 0 0.15 1 0 0 cm}#1\pdfliteral{1 0 -0.15 1 0 0 cm}}
\newcommand\mathbbsl[1]{\mathbb{\fakeslant{#1}}}

\DeclareMathAlphabet{\mathpzc}{OT1}{pzc}{m}{it}

\numberwithin{equation}{section}								

\begin{document}
\begin{flushright}
	MPP-2022-100
\end{flushright}
	$$\quad  $$
	{\LARGE\center \bfseries  Poisson-Lie T-duality defects\\ and target space fusion  \par } 
	\vspace{10pt}
	\begin{center}
	{\large\center Saskia Demulder and Thomas Raml}
	\end{center}
	\begin{center}
	{\small
		\textit{Max-Planck-Institut f\"ur Physik (Werner Heisenberg Institut),\\ F\"ohringer Ring 6, 80805 M\"unchen, Germany}	\\
			\vspace{1pt}
	E-mail: \texttt{\{sademuld,raml\}@mpp.mpg.de}}
		\end{center}
	\vspace{15pt}
	\noindent

	{\sc  Abstract.}  Topological defects have long been known to encode symmetries and dualities between physical systems. In the context of string theory, defects have been intensively studied at the level of the worldsheet. Although marked by a number of pioneering milestones, the target space picture of defects is much less understood. In this paper, we show, at the level of the target space, that Poisson-Lie T-duality can be encoded as a topological defect. With this result at hand, we can postulate the kernel capturing the Fourier-Mukai transform associated to the action of Poisson-Lie T-duality on the RR-sector. Topological defects have the remarkable property that they can be fused together or, alternatively, with worldsheet boundary conditions. We study how fusion of the proposed generalised T-duality topological defect consistently leads to the known  duality transformations for boundary conditions. Finally, taking a step back from generalised T-duality, we tackle the general problem of understanding the effect of fusion at the level of the target space. We propose to use the framework of Dirac geometry and formulate the fusion of topological defects and D-branes in this language.

	\tableofcontents
	
%%%%%%%%%%%%%%%%%%%%%%%%%%%%%%%%%%%%%%%%%%%
%%%%%%%%%%%     Introduction     %%%%%%%%%%	
%%%%%%%%%%%%%%%%%%%%%%%%%%%%%%%%%%%%%%%%%%%	
\section{Introduction}
Topological defects are known to offer a way to encode symmetries and dualities of a physical system. Studying symmetry and duality-encoding defects opens an alternative path to explore and understand symmetries  in a universal way, see \cite{Bachas:2012bj} for many references on the subject.  Topological defects have been extensively studied as objects living on the worldsheet of conformal field theories, see e.g. \cite{Bachas:2001vj,Brunner:2020miu,Oshikawa:1996ww}.  The monicker ``topological'' refers to their distinguishing feature: the defect line on the worldsheet can be deformed continuously. Indeed, the defining property of a topological defect is that the energy-momentum tensors of the models on both sides continuously connect at the location of the defect. As a consequence, since the energy-momentum tensor is the generator of diffeomorphisms, topological defects have the remarkable property that they can be moved freely on the worldsheet and superimposed either with another defect or a boundary condition on the worldsheet. This composition\footnote{In fact, in the conformal field theories fusion at the level of the corresponding operators corresponds to composing the associated defect operators, see e.g. \cite{Frohlich:2006ch} for additional details.} operation of topological defects (and boundary conditions) is known as ``fusion'' \cite{Bachas:2007td,Frohlich:2004ef}. The fusion algebra induced by topological defects is well-understood as the tensor product of bimodules in the representation category of the chiral symmetry algebra of the CFT.

In this article, we will however consider a \textit{target space} picture of topological defects. We will specialise to topological defects dividing two theories that are captured by non-linear sigma-model actions. We thus have two maps $X:\Sigma\rightarrow M$ and $\tilde X:\tilde \Sigma\rightarrow \tilde M$ from the worldsheets $\Sigma$ and $\tilde \Sigma$, that we will take to be separated by a defect line $D$, to the target spaces $M$ and $\tilde M$. As mentioned earlier, topological defects are objects living on the worldsheets where they can be moved and fused with other defects or boundary conditions on the worldsheet. Whatever happens on the worldsheet should translate to an object or operation on the target space. A line defect on the worldsheet is thus mapped by the product map $\mathbbsl X=X\times \tilde X$ to a submanifold $\mathcal Y$ of the product space $M\times \tilde M$. As is the case for D-branes, fields in the bulk couple to the defect via a line bundle determined by a connection living on the defect and the associated field strength $\mathcal F$. Demanding the defect line to be topological then translates into constrains on the possible pairs $(\mathcal Y,\mathcal F)$ on $M\times \tilde M$. 

The target space picture of topological defects was first considered in the context of WZW models in \cite{Fuchs:2007fw}, see also \cite{Runkel:2008gr}, and later in the context of generalised geometry in \cite{Kapustin:2010zc}. In \cite{Fuchs:2007fw}, by computing the scattering of closed string in a WZW model with D-branes, so-called biconjugacy classes were identified as the submanifolds $\mathcal Y$ realising the target space picture of the topological line defect on the worldsheet. Due to the similarity with the D-branes in WZW models, these topological defects were dubbed ``bibranes''. In what follows we will also at times use the term ``bibrane'' for any pair $(\mathcal Y,\mathcal F)$ defining a topological defect on the target space of a non-linear sigma-model.

Although it is considered a general fact that dualities and symmetries can be encoded by topological defects, it is not clear how much the notion of duality or symmetry can be stretched or weakened before this common lore breaks down. In string theory, Abelian T-duality is a prime example of a duality that can be encoded by a topological defect, see \cite{Frohlich:2006ch} for a worldsheet perspective and \cite{Sarkissian:2008dq,Kapustin:2003sg} for a target space derivation. Although very different in nature, non-Abelian T-duality \cite{Gevorgyan:2013xka}, was also shown to admit a description as a topological defect living on the target space of the dual models.

More recently, a more general form of T-duality has gained more and more attention. This duality takes a step back by no longer hinging on the existence of isometric directions in the background, instead demanding a weaker condition known as Poisson-Lie symmetry\cite{Klimcik:1995ux,Klimcik:1995dy}. Although featuring many similarities with Abelian T-duality, it still remains unclear if Poisson-Lie models can be considered as a symmetry between conformal systems. Indeed Poisson-Lie symmetric models generically do not produce solution to the conventional supergravity equations but a weaker set of equations, dubbed modified or generalised supergravity \cite{Arutyunov:2015qva}. This new set of equations still enjoys scale invariance, but is however likely to be plagued by a Weyl anomaly\footnote{A proposal to counter this problem involves considering the contribution of a modified Fradkin-Tseytlin term, see \cite{Wulff:2018aku,Sakamoto:2017wor,Fernandez-Melgarejo:2018wpg,Muck:2019pwj}. Note also that this problem can be circumvented by constructing the backgrounds using supergroups, where a non-unimodular r-matrix can always be chosen such that the background solves the un	modified supergravity equations \cite{Hoare:2018ebg,Hoare:2018ngg}.}. In addition, whilst conventional T-duality is a \textit{quantum} symmetry of string theory, it is still unclear if Poisson-Lie T-duality can be lifted to a quantum duality. Some initial steps towards  addressing this questions for the Poisson-Lie case were made in \cite{Tyurin:1995bu,Alekseev:1995ym}. The quantum aspects of Poisson-Lie symmetry have also been addressed more recently in \cite{Hassler:2020tvz,Hassler:2020wnp}. These observations raise the question of whether Poisson-Lie T-duality is ``enough of a duality'' to be described by a topological defect. In what follows, we will answer this question by the affirmative and we will unravel their properties under fusion. Let us first summarise the novel results presented in this article:

%%%%%%%%%%%%%%%%%%%%%%%%%%%%%%%%%%%%%
\subsubsection*{Summary of the results and overview}
\begin{itemize}
	\item[$\triangleright$] We show, by direct construction, that Poisson-Lie T-duality can be represented, by using a target space formulation, as a topological defect. The line-bundle of this defect is controlled by the so-called Semenov-Tian-Shansky symplectic form on the Drinfel'd double associated to the Poisson-Lie T-dual models. We establish the existence of a Poisson-Lie T-duality inducing topological defect when the model includes spectator fields.
	\item[$\triangleright$] Using the previous result, we provide further evidence that the Fourier-Mukai transform for Poisson-Lie T-duality has a kernel that is essentially captured by the Semenov-Tian-Shansky symplectic form of the Drinfel'd double, as was proposed recently in \cite{Arvanitakis:2021lwo}.
	\item[$\triangleright$] We show how the known transformation rules for the gluing conditions of open strings under generalised T-duality can be derived by fusing the Poisson-Lie T-duality defect with a boundary condition.
	\item[$\triangleright$] In the last section,  no longer specialising to Poisson-Lie symmetric backgrounds, we investigate the result of fusing topological defects and boundary conditions in terms of the associated target space data. To do so, we first underline the universal description of D-branes and defects in terms of Dirac geometry and subsequently exploit this observation to formulate a notion of fusion at the level of target space. 
\end{itemize}

\noindent
In section \ref{sec:PL_top}, we first show how to construct Poisson-Lie T-duality as a topological defect. This can be established in two ways, either in terms of the Lagrangian and its equations of motion or by lifting the discussion to the associated product space. We will establish the existence of the duality defect in both formalisms. Whilst the former is the most prevalent approach for Lagrangian theories, the latter method will act as a stepping stone for the later discussion of fusion. Section \ref{sec:fusion} discusses the process of fusion of topological defects at the level of the target space. With the generalised T-duality topological defect at hand, in section \ref{sec:BC_gluing}, we show how fusing the generalised T-duality defect with boundary conditions characterised by gluing conditions yields the known duality transformations. Finally, in section \ref{sec:fusion_via_DS}, we will take a step back from generalised T-duality. There, we consider the problem of understanding the well-established notion of worldsheet fusion of topological defects on the target space. We combine the expectation that the fusion process should mirror a Fourier-Mukai transformation together with the insight that Dirac structures are the fitting objects to describe the worldvolume of D-branes and topological defects. We then propose a fusion operation that uses the language of generalised and Dirac geometry.
In section \ref{sec:discussion}, we draw some final conclusions and future perspectives. We complement the discussion in the main text with a number of appendices.

%%%%%%%%%%%%%%%%%%%%%%%%%%%%%%%%%%%%%%%%%%%
%%%%%%%%%%%%%%%%%%%%%%%%%%%%%%%%%%%%%%%%%%%
%%%%%%%%%%%%%%%%%%%%%%%%%%%%%%%%%%%%%%%%%%%
\section{Poisson-Lie T-duality as a topological defect}\label{sec:PL_top}
In this section, we establish Poisson-Lie T-duality as a space-filling topological defect carrying a line bundle with curvature given by the so-called Semenov-Tian-Shansky (STS) symplectic structure, both with and without spectator directions. After showing this first at the Lagrangian level, and anticipating on our discussion on fusion in the second part of this paper, we will also show this fact directly at the level of the product space. This will allow us, following \cite{Kapustin:2010zc}, to later formulate the defects in terms of Dirac structures.   

Before introducing the topological defect relevant to Poisson-Lie T-duality, we first review some necessary facts on this generalised notion of the T-duality as well its associated algebraic objects. See \cite{Klimcik:1995dy,Klimcik:1995ux} or the reviews \cite{Klimcik:1995jns,Demulder:2019bha,Klimcik:2021bjy} for a more detailed and thorough discussion of Poisson-Lie T-duality. In what follows, we will consider sigma-models that are Poisson-Lie symmetric. This ``symmetry'' can be seen as a weaker requirement than demanding to have Abelian or non-Abelian directions in Abelian, resp. non-Abelian, T-duality. The general form of a sigma-model admitting Poisson-Lie symmetry is as follows. We have a map $g(X):\Sigma\rightarrow M$, where we assume a Lie group $\mathcal G$ acts on $M$, together with the action \cite{Klimcik:1995ux}
\begin{align}\label{eq:PLaction}
	S_\mathrm{PL}=\int_\Sigma L^a  (E_0^{-1}+\Pi[g])^{-1}_{ab}L^b\,,
\end{align}
where we have denoted by $X$ a set of coordinates on a local patch of $M$, the left-invariant Maurer-Cartan form are denoted by $L = L^aT_a=l^a_i \mathrm d X^i T_a = g^{-1}  \mathrm d g$ for $g\in M$ and $T_a$ are a set of generators for the Lie algebra  $\mathrm{Lie}(\mathcal G)=\frak g$ (and likewise for the tilde variables) with $a=1,\cdots,\mathrm{dim}\, \mathcal G$, $E_0$ is a $\mathrm{dim}\,M\times \mathrm{dim}\,M$-dimensional constant matrix and $\Pi[g]$ is a Poisson-Lie structure obtained by solving the (modified) classical Yang-Baxter equation for the Lie algebra $\mathrm{Lie}(\mathcal G)=\frak g$. The underlying reason to why such models naturally generalise the conventional form of T-duality is that its underlying symmetry group is a Poisson-Lie group, that comes with a natural notion of ``dual object'' baked in. Indeed, the Poisson-structure $\Pi[g]$ associated to $\mathcal G$ does not only define a (multiplicative) Poisson structure on $\mathcal G$ it determines in fact  a second Lie bracket on the algebra $\frak g$. Denote by $\tilde{\frak{g}}$ the algebra defined by this second bracket on the vector space $\frak g$. The associated Lie group $\tilde{\mathcal G}$ also admits a Poisson-Lie structure $\tilde \Pi$. The relation is symmetric: the Poisson-Lie dual of $\tilde{\frak{g}}$ is again ${\frak{g}}$. One can write an action with target space $\tilde{ \mathcal G}$; the Poisson-Lie T-dual action, 
\begin{align}\label{eq:dualPLaction}
	\tilde S_\mathrm{PL}=\int_{\tilde \Sigma}  \tilde L_a   [( E_0+\tilde \Pi[\tilde g])^{-1}]^{ab} \tilde L_b\,,
\end{align}
where $\tilde \Pi[\tilde g]$ is the Poisson-Lie structure on the Lie group $\tilde{\mathcal G}$ associated to the double $\mathbbsl D=\mathcal G\tilde{\mathcal G }$ and $\tilde l$ are now the components of the left-invariant Maurer-Cartan form $\tilde L=\tilde l_{ai} \mathrm d \tilde X^i \tilde T^a$with respect to an element $\tilde g\in \tilde{\mathcal G}$. Remarkably, as was shown in \cite{Klimcik:1995ux,Klimcik:1995dy}, the actions in eqs. \eqref{eq:PLaction} and \eqref{eq:dualPLaction} naturally generalise the notion of T-dual models by taking $\mathbbsl{D}=U(1)^{2\dim \mathcal G}$ for Abelian T-duality and $\mathbbsl{D}=\mathcal G\times U(1)^{\dim \mathcal G}$ for non-Abelian T-duality.

There is a last but crucial fact left to introduce. The direct sum of the Lie algebra $\frak g$ with its Poisson-Lie T-dual algebra $\tilde{\frak{g}}$ leads to a so-called  Drinfel'd double $\frak d=\frak g\oplus \tilde{\frak{g}}$. See appendix \ref{app:STS} for definitions and details. An element $\mathbbsl{g}$ of the exponentiation $\mathbbsl{D}$ of this algebra, which is also called the Drinfel'd double of $(\mathcal G,\Pi)$, has by construction, locally, two possible decompositions
\begin{align}
\mathbbsl{g} = \tilde{h} g \quad \text{or} \quad \mathbbsl{g} = h \tilde{g}\,,\quad \text{with}\;\,\,g,h\in \mathcal G\,,\ \tilde g,\tilde h\in \tilde{ \mathcal G}\,,
\end{align}
where  the Drinfel'd double is assumed to be perfect, such that these decompositions are unique.
In addition, this Lie group $\mathbbsl D$ can be naturally endowed with a symplectic structure, called the Semenov-Tian-Shansky (STS) symplectic form  \cite{semenov2008integrable,Alekseev:1993qs}
\begin{align}\label{eq:STS}
  \omega_\mathrm{STS}  &=\tfrac{1}{2}( r^a(g) \wedge \tilde{l}_a(\tilde{h}) - \tilde{r}_a(\tilde{g}) \wedge l^a(h))\,,
\end{align} 
where we have used the components of the right-invariant Maurer-Cartan form $R = r^a_i \mathrm d X^i T_a =\mathrm d gg^{-1}$ for $g\in \mathcal G$.  The Drinfel'd double together with the Semenov-Tian-Shansky symplectic form will play a crucial role in what follows.

%%%%%%%%%%%%%%%%%%%%%%%%%%%%%%%%%%%%%%%%%%%
%%%%%%%%%%%%%%%%%%%%%%%%%%%%%%%%%%%%%%%%%%%
\subsection{Sigma-model approach}\label{sec:sigma_model_approach}
At the level of the target space, a topological defect can be described by a total Lagrangian involving the sigma-model actions of the neighbouring theories together with a term specifying the coupling of the line bundle on the defect worldvolume to the bulk fields. We thus consider a pair of target space maps $X:\Sigma\rightarrow M$ and $\tilde X:\tilde\Sigma\rightarrow \tilde M$. We glue these two theories together by combining the worldsheets $\Sigma \cup \tilde \Sigma$ along a defect line $D$, located at $\sigma=0$, so that the target space map $X(\tau,\sigma)$ is now defined  for $\sigma \geq 0$ and $\tilde X(\tau,\sigma)$  for $\sigma \leq 0$. It is thus convenient to define the product target space map
\begin{align}\label{eq:product_target_space_map}
	\mathbbsl{X}=X\times \tilde X\,:\,\Sigma \cup \tilde \Sigma\rightarrow M\times \tilde M:(\tau,\sigma)\mapsto\mathbbsl X^I=(X^i,\tilde X^i)\,.
\end{align}
The pushforward of the product map $\mathbbsl{X}$ restricted to the defect line $D$ will then span a submanifold $\mathcal Y$ in product space $M\times \tilde M$. In analogy with D-branes, the bulk fields will couple to a line bundle whose pullbacked connection  $\mathbbsl A$ lives on the worldvolume of the submanifold $\mathcal Y$ corresponding to the defect. The curvature of the line bundle connection will be denoted by $\mathcal F=\mathrm d \mathbbsl A$. That is, from the perspective of the target space, a defect between two theories with target spaces $M$ and $\tilde M$ is a pair $(\mathcal Y,\mathcal F)$ consisting of a submanifold $\mathcal Y$ in the product space $M\times \tilde M$ together with a two-form $\mathcal F$. We thus have a total worldsheet action of the form \cite{Fuchs:2007fw,Sarkissian:2008dq}
\begin{gather}\label{eq:tot_action}
\begin{aligned}
	S&=\frac{1}{2}\int_\Sigma\mathrm d^2\sigma\,\left(G(X)_{ij}\partial_\mu X^i\partial^\mu X^i+ B(X)_{ij}\epsilon^{\mu\nu}\partial_\mu X^i\partial_\nu X^j\right)\\
	&\qquad +\frac{1}{2}\int_{\tilde\Sigma}\mathrm d^2\sigma\,\left(\tilde G(X)_{ij}\partial_\mu\tilde X^i\partial^\mu\tilde X^i+ \tilde B(X)_{ij}\epsilon^{\mu\nu}\partial_\mu\tilde X^i\partial_\nu\tilde X^j\right)
	+\int_D\mathrm d\tau\,\mathbbsl A_I(\mathbbsl X)\partial_\tau\mathbbsl X^I\,.
\end{aligned}
\end{gather}
 Locally, the three-form fluxes $H$ and $\tilde H$ in each model are identified with the exterior derivative of the Kalb-Ramond two-forms $B_{ij}$ and $\tilde B_{ij}$, i.e. $H=\mathrm d B$ and $\tilde H=\mathrm d \tilde B$. The geometric data associated to the defect can be encoded into a two-form field $F$ defined on the topological defect by $F=-(p^\ast B-\tilde p^\ast \tilde B+\mathcal F)$, where the overall minus sign is chosen for later convenience and where $p$ and $\tilde p$ are the projections from the product space onto $M$, respectively $\tilde M$ and the Kalb-Ramond fields are evaluated at the location of the defect. Note that by definition we have that  the corresponding exterior product is the difference of the three-form fluxes of the theories on the left and on the right $\mathrm d F =H-\tilde H$, when evaluated at the location of the defect.

It is clear that not all pairs of submanifolds and line bundle curvatures $(\mathcal Y,\mathcal F)$ realise the worldvolume of a topological defect. Indeed the topological property given in eq. \eqref{eq:top_def_cond}, that requires the energy momentum tensors of the two neighbouring models to continuously connect along the defect line, is highly constraining. The energy momentum tensor of a sigma-model is given by
\begin{align*}
	T_{\mu\nu}=G_{ij}(X) \partial_\mu X^i\partial_\nu X^j-\frac{1}{2}g_{\mu\nu}g^{\gamma\delta} G_{ij}(X)\partial_\gamma X^i\partial_\delta X^j\,,
\end{align*}
where $g_{\mu\nu}$ is the worldsheet metric, or in terms of the coordinates $z=\tau +\sigma$ and $\bar z=\tau -\sigma$ one has
\begin{align}\label{eq:exprT}
T=G_{ij}\partial X^i\partial X^j\,,\quad \bar T=G_{ij}\bar \partial X^i\bar \partial X^j\,,
\end{align}
were the derivatives are given by  $\partial = \frac{1}{2}(\partial_\tau + \partial_\sigma)$ and  $\bar{\partial} = \frac{1}{2}(\partial_\tau - \partial_\sigma)$.
A defect is called topological when it continuously connects holomorphic and anti-holomorphic components of the left- and right energy-momentum tensors\footnote{A weaker notion of defect is that of conformal defects for which we require the conservation of the worldsheet energy. This translates into the requirement that the off-diagonal components of the stress-energy tensors glue continuously
$T^1{}_0-\hat T^1{}_0=0$,
at the location of the defect $D$. Topological defects are a stronger concept, by additionally demanding that the worldsheet momentum is also conserved. That is guaranteed when at the defect the diagonal components of the stress-energy tensor glue continuously $
	T^1{}_1-\hat T^1{}_1=0\,.$
Topological defects in contrast to conformal defects can be, as long as the defect does not cross the location of a local operator insertion, deformed smoothly without affecting the values of correlators. For conformal defects, one has to be more careful as the fusion operation might lead to singularities and corresponding divergences.\label{ftn:conformal_defects}} 
\begin{align}\label{eq:top_def_cond}
	T=\tilde T\,,\qquad \bar{T}=\bar{\tilde{T}}\,,
\end{align}
at the location of the defect. The constraint in eq. \eqref{eq:top_def_cond} has striking implications. Indeed, the topological defect condition in eq. \eqref{eq:top_def_cond} implies that, since the energy-momentum is the generator of diffeomorphisms, the total system given by the action in eq. \eqref{eq:tot_action}, remains invariant under continuous deformations of the defect at values $\sigma\neq 0$. This observation leads to one of the most striking features of this class of defects: the process of fusion. Topological defects can  be continuously deformed away from their initial location on the worldsheet. When two defects are superimposed, they fuse into one or multiple new defects. In a like manner, when the sigma-models describe open strings, a topological defect can be moved to the boundary of the worldsheet and fused with the boundary conditions resulting in a new boundary condition. On a general basis, invertible\footnote{A topological defect is called invertible if there exists a second defect such that their composition or fusion yield the invisible defect. In conformal field theories, where defects can be associated to operators, invertibility means a trivial null space. As a result theories separated by an invertible defect share the same spectrum. See e.g. \cite{Davydov:2010rm} for a more rigorous treatment.} topological defects are expected to implement dualities between theories.  
Indeed, in \cite{Sarkissian:2008dq,Gevorgyan:2013xka}, it was shown how, at the level of the target space formulation, particular topological defects can encode T-duality and even non-Abelian T-duality. In the following section we will prove that one can also construct a topological defect encoding the most general form of T-duality on group manifolds; Poisson-Lie T-duality. 

%%%%%%%%%%%%%%%%%%%%%%%%%%%%%%%%%%%%%%%%%%%	
\subsubsection{Poisson-Lie T-duality topological defect }\label{sec:Lagr_def_no_spect}
To show that the pair $(\mathbbsl D,\omega_\mathrm{STS})$ realises the worldvolume of a topological defect encoding Poisson-Lie T-duality, the starting point is the total action in eq. \eqref{eq:tot_action}, i.e.
\begin{align}\label{eq:tot_act_compact}
S_\mathrm{tot} = S+\tilde S  +S_\mathrm{defect}=2\int_\Sigma L^a  E_{ab} L^b + 2 \int_{\tilde \Sigma}   \tilde  L_a \tilde E^{ab}
	\tilde L_b  + \int_D \mathbbsl{X}^\ast \mathbbsl{A}\,,
\end{align}
where $E=G+B$ is the background field consisting of metric $G$ and $B$ the antisymmetric Kalb-Ramond field and likewise for the tilde variables.  The left-invariant Maurer-Cartan form is denoted, as above, by $L = l^a_i \mathrm d X^i T_a = g^{-1}  \mathrm d g$ for $g\in M$ and $T_a$ are a set of generators for the Lie algebra  $\mathrm{Lie}(\mathcal G)=\frak g$ (and likewise for the tilde variables) with $a=1,\cdots,\mathrm{dim}\, \mathcal G$. The map $\mathbbsl{X}=X\times \tilde X|_D$ is the product target space map defined in eq. \eqref{eq:product_target_space_map}. The first action, given by $S$, is taken to be a  Poisson-Lie symmetric model on a Lie group manifold, as given in eq. \eqref{eq:PLaction} and the second action $\tilde S$ is left unspecified. We will make the Ansatz that the defect is given by the pair $(\mathbbsl D,\omega_{STS})$. That is we take $\mathrm d\mathbbsl A(X,\tilde X)=\mathcal F= \omega_\mathrm{STS}(X,\tilde X)$, where the Semenov-Tian-Shansky symplectic form was defined in eq. \eqref{eq:STS}. We will see that by picking a particular solution of the equation of motion in the presence of the defect, the background fields of the target space $\tilde{ \mathcal G}$ will be precisely the Poisson-Lie T-dual action $\tilde S_\mathrm{PL}$ given in eq. \eqref{eq:dualPLaction}. In addition, we show that this defect is topological.  

\noindent
First, it will be more convenient to use a rewriting of this symplectic form
\begin{align}\label{eq:STS_worked_out}
	2 \omega_{STS} = 2 l^b(g)  \wedge C_b^{\ e}\tilde{l}_e(\tilde{g}) +   l^b(g)  \wedge C_b^{\ m} \tilde{\Pi}_{me}[\tilde{g}] l^e(g)-\tilde{l}_b(\tilde{g})  \wedge \tilde{C}^b_{\ m} \Pi^{me}[g]  \tilde{l}_e(\tilde{g}) \,,
\end{align}
where we have introduced $C=(1-\tilde \Pi\Pi)$ and similarly for $\tilde C$ with the order of $\Pi, \tilde \Pi$ switched.
This expression for the STS symplectic form is derived in the appendix \ref{app:STS} or see also \cite{Arvanitakis:2021lwo} for an equivalent but slightly different form. Varying now the action with respect to $X^i$ and keeping only terms localized at $\sigma=0$ we get
\begin{align}
 \int_D \mathrm d \tau \Bigl. \Bigl( E_{ab}  l^b_{ j}  \bar{\partial} X^j - E_{ba} l^b_{j}\partial X^j \Bigr)\Bigr|_{\sigma=0} l^a_{\ k} \delta X^k = - \left. \delta_{X^i} \left( \int_D \mathbbsl{X}^{\ast} \mathbbsl{A} \right) \right|_{\sigma=0} \,,
\end{align}
and similarly for $\delta_{\tilde{X}^i}$.
Taking the variation of the boundary term
\begin{align*}
	\delta_{X^i} \int_D \mathbbsl{X}^\ast \mathbbsl{A} &=\int_D \mathrm d \tau \left(C_a^{\ b} \tilde{\Pi}_{bc}[\tilde{g}] l^c_j \partial_\tau X^j + C_a^{\ b} \tilde{l}_{bj} \partial_\tau \tilde{X}^j  \right) l^a_i \delta X^i\,.
\end{align*}
From these variations the following equations of motion on the defect line $\sigma=0$ follow:
\begin{gather}\label{eq:EOMs_at_defect}
\begin{aligned}
E_{b a} l^b_j \partial X^j  - E_{a b} l^b_j \bar{\partial}X^j &=   C_a^{\ b}\tilde{\Pi}_{bc}[\tilde{g}] l^c_j \partial_\tau X^j + C_a^{\ b} \tilde{l}_{bj} \partial_\tau \tilde{X}^j\,,\\
\tilde{E}^{b a} \tilde{l}_{b j} \partial \tilde{X}^j  - \tilde{E}^{a b} \tilde{l}_{b j} \bar{\partial}\tilde{X}^j &=  \tilde{C}^a_{\ b} \Pi^{bc}[g]  \tilde{l}_{cj} \partial_\tau \tilde{X}^j + l^b_{j} C_b^{\ a} \partial_\tau X^j \,.
\end{aligned}	
\end{gather}
To solve these equations of motion, we pick a particular solution which, as we will see in the next section, singles out the topological defect encoding Poisson-Lie T-duality. We set
\begin{gather}
\begin{aligned}\label{eq:Eab on defect}
&\left(E_{ab} + C_a^{\ c}\tilde{\Pi}_{cb}[\tilde{g}] \right) l^b_j \bar{\partial} X^j =  -   C_a^{\ b} \tilde{l}_{bj}  \bar{\partial}  \tilde{X}^j\,,\quad \left(E_{ba} -  C_a^{\ c} \tilde{\Pi}_{cb}[\tilde{g}] \right) l^b_j \partial X^j =   C_a^{\ b} \tilde{l}_{bj} \partial \tilde{X}^j\\
&\left(\tilde{E}^{ab} + \tilde{C}^a_{\ c}\Pi^{cb}[g] \right) \tilde{l}_{b j} \bar{\partial} \tilde{X}^j = -  C_b^{\ a} l^b_{j}\bar{\partial}  X^j\,,\quad \left(\tilde{E}^{ba} - \tilde{C}^a_{\ c}\Pi^{cb}[g]  \right) \tilde{l}_{b j} \partial \tilde{X}^j =    C_b^{\ a} l^b_{j}\partial  X^j\,.
\end{aligned}
\end{gather}
One can readily check that these relations, if satisfied, solve the equations of motion at the location of the defect line.
Let us now show that the relations in eq. \eqref{eq:Eab on defect} are nothing but the canonical relations for Poisson-Lie T-duality \cite{Sfetsos:1997pi} which we review in appendix \ref{app:Canonical_PL}. Indeed, one can rewrite the eqs. in eq. \eqref{eq:Eab on defect} as
\begin{gather}\label{eq:eom_before_consistency_compact}
\begin{aligned}
	\tilde C\bar L&=-\tilde C(E+C\tilde \Pi)^{-1}C\tilde L\,,\quad \tilde CL=\tilde C(E^T-C\tilde \Pi)^{-1}C\tilde L \,,\\
	 C\bar{\tilde{L}}&=- C(\tilde E+\tilde C \Pi)^{-1}\tilde C L\,,\quad  C\tilde L= C(\tilde E^T-\tilde C \Pi)^{-1}\tilde C L \,,
\end{aligned}
\end{gather}
where we have suppressed the indices for ease of notation.
Consistency of the equations in eq. \eqref{eq:eom_before_consistency_compact} with respect to those in eq. \eqref{eq:Eab on defect} forces us to identify 
\begin{align}\label{eq:top_def_cond_eoms}
\tilde{E}^{ab} + \tilde{C}^a_{\ c}\Pi^{cb}[g] \equiv  \tilde{C}^a_{\ d} \left( (E + C\tilde{\Pi}[\tilde{g}] )^{-1} \right)^{dc} C_c^{\ b}\,,
\end{align}
and similarly for the other equations.  These equations are however nothing but the canonical transformation rules for the background fields of Poisson-Lie T-dual models \cite{Sfetsos:1997pi} 
\begin{gather}\label{eq:canonical_PL} 
\begin{aligned}
		C^{-1}(E+C\tilde{\Pi})\tilde{C}^{-1}(\tilde{E}+\tilde{C}\Pi)=1\,,\qquad C^{-1}(E^T - C\tilde{\Pi})\tilde{C}^{-1}(\tilde{E}^T-\tilde{C}\Pi)=1\,.
\end{aligned}
\end{gather}
That is, we have shown that defect $(\mathcal Y,\mathcal F)=(\mathbbsl D,\omega_\mathrm{STS})$ encodes the Poisson-Lie T-duality transformation.

Having established that the action in eq. \eqref{eq:tot_act_compact} with sigma-models given by eqs. \eqref{eq:PLaction} and \eqref{eq:dualPLaction} and with one-form field defined via $\mathrm d \mathbbsl A=\omega_\mathrm{STS}$  describes a defect encoding Poisson-Lie T-duality, we now show that the defect is in fact topological. That is, the defect verifies the continuity of the energy-momentum tensors as give in eq. \eqref{eq:top_def_cond} across the defect location. The holomorphic part of the energy-momentum tensors for the left  Poisson-Lie models, after using the relation in eq. \eqref{eq:Eab on defect} in the expression given in eq. \eqref{eq:exprT}, leads to 
\begin{align*}
	T&= C_b^{\ c} \tilde{\Pi}_{ca} l^a_i l^b_j \partial X^i \partial X^j +  C^{\ a}_{b}  \tilde{l}_{ai} l^b_j \partial \tilde{X}^i \partial X^j \,,
\end{align*}
whilst the energy-momentum tensor for the dual right model takes on the form
\begin{align*}
\tilde{T}&=   \tilde{C}^b_{\ c} \Pi^{ca} \tilde{l}_{ai} \tilde{l}_{bj} \partial \tilde{X}^i \partial \tilde{X}^j + C_b^{\ a}l^b_j \tilde{l}_{ai} \partial X^j \partial \tilde{X}^i\,.
\end{align*}
Taking their difference we see that the last term from each energy momentum tensor cancel, leading to 
\begin{equation}
T-\tilde{T}= \Pi^{ac} C_c^{\ b}   \tilde{l}_{ai} \tilde{l}_{bj} \partial \tilde{X}^i \partial \tilde{X}^j - \tilde{\Pi}_{ac} \tilde{C}^c_{\ b} l^a_i l^b_j \partial X^i \partial X^j  =0\,,
\end{equation}
where the left-over terms vanish identically by virtue of $\Pi C$ and $\tilde\Pi\tilde C$ being antisymmetric\footnote{This can be shown using the explicit form of $C, \tilde{C}$ in terms of adjoint actions given in the appendix \ref{app:STS}.}. One can verify in a very similar fashion that the condition on the antiholomorphic part of the energy-momentum tensors on both theories is verified, showing that the defect defined by the total system defined in eq. \eqref{eq:top_def_cond_eoms} with field strength $\mathrm d  \mathbbsl A=\omega_\mathrm{STS}$ given by the Semenov-Tian-Shansky symplectic structure is indeed topological.

%%%%%%%%%%%%%%%%%%%%%%%%%%%%%%%%%%%%%%%%%%%	
\subsubsection{Other topological defects}
In \cite{Kapustin:2009av} and \cite{Bachas:2012bj}, the relation between the fusion algebra of topological defects in conformal field theories and the duality group $O(d,d)$ (and its semi-group extension) was studied. The group $O(d,d)$ is not only generated by (factorised) T-dualities but also linear transformations and $B$-field transformations. As later highlighted in \cite{Sarkissian:2008dq,Elitzur:2013ut}, one can indeed also realise the latter two, i.e the $B$-field transformations and linear transformations, using the target space formulation of defects.  $B$-field transformations between the background field $E=G+B$ on a target space $M$ and the background field $\tilde E$ on $\tilde M=\mathcal G$, are associated with a ``diagonal defect'' determined by the choice $X^i=\tilde X^i$ in the product space $\mathcal G\times \mathcal G$ with the two-form field $B$. On can show that this defect is again topological, we will not repeat the argument in detail, which can be found in \cite{Sarkissian:2008dq,Elitzur:2013ut}. What is relevant to remember for what is to come, is that the worldvolume of this defect is realised by a pair $(\mathcal{G}_\mathrm{diag},B)$, where $\mathcal{G}_\mathrm{diag}$ is a copy of $\mathcal G$ diagonally embedded in the product space $\mathcal G\times \mathcal G$. 

Finally, let us remark that for a given Poisson-Lie symmetric target space $\mathcal G$ there are, by construction, as many Poisson-Lie T-duality defects as there are solutions of the (modified) classical Yang-Baxter equation for its algebra $\frak g$, or equivalently so-called Manin triples associated with $\frak g$. To see this, we need yet another equivalent expression for the Semenov-Tian-Shansky symplectic form. For each solution $r$ of the modified classical Yang-Baxter equation for the Lie algebra $\frak g$, we obtain a different dual algebra $\tilde{\frak{g}}$ or Drinfel'd double $\frak d=\frak g\oplus\tilde{\frak{g}}$\footnote{Technically this means that we are looking at ``split'' type classical modified Yang-Baxter equation. Indeed, solutions to the non-split classical modified Yang-Baxter equation for a fixed Lie algebra $\frak g$ is essentially unique. We will not discuss solutions to the homogeneous classical Yang-Baxter equation. More details and definitions can be found in \cite{semenov2008integrable,Vicedo:2015pna}.}. It turns out that the Drinfel'd double $\frak d$ is itself also a Poisson-Lie group with r-matrix $r_\mathfrak{d}$ given by the expression 
\begin{align*}
	r_\frak{d}=p_{\frak g}- p_{\tilde{\frak{g}}}\,,
\end{align*}
where $p_\frak{g}$ is the projection into $\frak g$ seen as diagonally embedded into $\frak d$ and $p_{\tilde{\frak{g}}}$ is the projection into the Poisson-Lie dual algebra $\tilde{\frak{g}}$. In terms of the r-matrix $r_\frak{d}$ of the double, the Semenov-Tian-Shansky symplectic form, or equivalently the corresponding Poisson structure, can be expressed by, see e.g. \cite{semenov2008integrable}:
\begin{align}\label{eq:STS_rd}
	\{\,\cdot\,,\,\cdot\,\}_\mathrm{STS}=\tfrac{1}{2}\langle (r_\frak{d}-\mathrm{Ad}_{\mathbbsl{g}^{-1}}r_\frak{d}\mathrm{Ad}_{\mathbbsl{g}})\nabla\cdot,\nabla\cdot\rangle\,,
\end{align}
where the adjoint action are taken with respect to an element $\mathbbsl g\in \mathbbsl D$ of the double and $\nabla$ denotes the left gradient derivative, i.e. for a function $f\in C^\infty (\mathcal G)$ it takes the form $\langle \nabla f(x),X\rangle=\tfrac{\mathrm d}{\mathrm d t} f(\exp(tX)x)|_{t=0}$, for $X\in \frak g$.
That is, for each solution of the (modified) classical Yang-Baxter associated to the Lie algebra $\frak g$ we also have a different Drinfel'd double and associated Semenov-Tian-Shansky symplectic form is given in eq. \eqref{eq:STS_rd}.
%%%%%%%%%%%%%%%%%%%%%%%%%%%%%%%%%%%%%%%%%%%	
\subsubsection{Including spectators }\label{sec:Lagr_def_w_spect}
When considering spectator fields, the Lie group $\mathcal G$ acts freely, but no longer transitively on the target space $M$. We assume thus that the target space can be identified with a space of the form $M=\mathcal G\times N$ and is described by coordinates $X^m=(X^i, Y^\mu)$ with $i=1,\cdots, \dim \mathcal G=n$ and $N$ is the spectator manifold with coordinates $Y$. In this section, we consider a topological defect lying between two Poisson-Lie T-dual models but including the freedom of having spectator directions. We thus start from the same formal action
\begin{align}\label{eq:tot_act_spect}
	S_\mathrm{tot,spect}=S+\tilde S + S_\mathrm{defect}\,,
\end{align}
where we will first assume that the model on the left of the defect, given by the action $S$ is Poisson-Lie symmetric with spectator fields. 
In particular, this means that the first action is a non-linear sigma-model  taking the form
\begin{align}\label{eq:act_PL_spects}
	S &=2\int_\Sigma\mathrm d ^2 \sigma \, \begin{pmatrix}
		l^a_i \partial X^i &\partial Y^\mu 
	\end{pmatrix}\begin{pmatrix}
		E_{ab}&E_{a \nu}\\
		E_{\mu b} & E_{\mu\nu}
	\end{pmatrix}\begin{pmatrix}
		l_i^b\bar\partial X^i &\bar\partial  Y^\nu 
	\end{pmatrix}\,.
\end{align}
The form of components of the background field $E_{mn}$ are fixed by imposing Poisson-Lie symmetry and thus has to satisfy the condition 
\begin{align*}
	 L_{l_a}E_{mn}=-\tilde f^{bc}{}_a E_{mp}l^p{}_b l^q{}_c E_{qn}\,,
\end{align*}
where $\tilde f^{ab}{}_c$ are the structure constants of the algebra of the Lie group manifold $\tilde{\mathcal G}$ and one can determine the total background field $E_{mn}$ as \cite{Klimcik:1995ux,Klimcik:1995dy,Sfetsos:1997pi}
\begin{gather}
\begin{aligned}\label{eq:gen_E_PL_spect}
\begin{matrix*}[l]
E_{a\mu} = E_{ab} ((E_0)^{-1})^{bc} F_{c\mu}\,,  & E_{\mu \nu} = F_{\mu \nu} - F_{\mu a}\Pi^{ab}E_{bc}((E_0)^{-1})^{cd}F_{d \nu}\,, \\
\tilde{E}^{a}_{\ \mu} = \tilde{E}^{ab}F_{b \mu}\,, & \tilde{E}_{\mu \nu} = F_{\mu \nu} - F_{\mu a}\tilde{E}^{ab}F_{b \nu}\,.
\end{matrix*}
\end{aligned}
\end{gather}
Here $F_{\mu \nu}, F_{a \mu}$ are arbitrary matrices depending on the choice of  background configuration chosen. These reduce to the conventional Poisson-Lie symmetric backgrounds upon setting the spectators to zero. Crucially, the Poisson structure $\Pi$ of this general solution for the background field $E$ in presence of spectator fields, only depends on the coordinates $X$ on $\mathcal G$ and not on the spectator directions $Y$. The matrix  $(E_0)_{mn}=(E_0)_{mn}(Y)$ of coupling constant can however  have a $Y$-dependent, but no dependence on the coordinates of the manifold $\mathcal G$ or $\tilde{\mathcal G}$.

 The last term $S_\mathrm{defect}$ in the total action in eq. \eqref{eq:tot_act_spect} accounts for the one-form $\mathbbsl A$ on the defect line $D\subset \Sigma$, contributing with the usual term
 \begin{align*}
 	S_\mathrm{defect}=\int_D \mathbbsl X^\ast \mathbbsl{A}\,,
 \end{align*}
where $\mathbbsl X$ is the product map defined in eq. \eqref{eq:product_target_space_map} from the union of the worldsheet $\Sigma\cup \tilde \Sigma$ to the product space $\mathcal G\times \tilde{\mathcal G}\times N$. Note that we do not take two copies of the spectator manifold $N$. We will show that taking the simplest Ansatz, that is taking $\mathrm d\mathbbsl A(X,\tilde X)=\omega_\mathrm{STS}(X,\tilde X)$, will give a topological defect that encoded Poisson-Lie T-duality \textit{with} spectator fields.  We will see that taking again the STS symplectic structure to be the curvature of the line bundle of the topological defect takes care of the presence of the spectator fields, and will fix the model described by the action $\tilde S$ on the right to be the Poisson-Lie T-dual action, taking on the same form as the Poisson-Lie action $S$ in eq. \eqref{eq:act_PL_spects}, but defined by a set of background fields $\tilde E$ and (left-invariant) Maurer-Cartan forms $\tilde L=\tilde l_{ai}\tilde T^a\mathrm d X^i$ depending on the coordinates $(\tilde X,Y)$, sharing the same $Y$-direction (the spectator fields).

Varying the total action given in eq. \eqref{eq:tot_act_spect} and using again the alternative expression for the STS form given in eq. \eqref{eq:STS_worked_out}, the equations with spectator directions become
\begin{align*}
E_{b a} l^b_j \partial X^j + E_{\mu a} \partial Y^\mu - E_{a b} l^b_j \bar{\partial}X^j - E_{a \mu}  \bar{\partial} Y^\mu =   C_a^{\ b}\tilde{\Pi}_{bc}[\tilde{g}] l^c_j \partial_\tau X^j + C_a^{\ b} \tilde{l}_{bj} \partial_\tau \tilde{X}^j\,,
\end{align*}
and similarly for the tilde variables
\begin{align*}
\tilde{E}^{b a} \tilde{l}_{b j} \partial \tilde{X}^j + \tilde{E}^{\mu a}  \partial Y^\mu - \tilde{E}^{a b} \tilde{l}_{b j} \bar{\partial}\tilde{X}^j - \tilde{E}^{\mu b}  \bar{\partial} Y^\mu =  \tilde{C}^a_{\ b} \Pi^{bc}[g]  \tilde{l}_{cj} \partial_\tau \tilde{X}^j + l^b_{j} C_b^{\ a} \partial_\tau X^j \,.
\end{align*}
Finally varying with respect to the spectators $Y^\mu $ we now get a third equation
\begin{align}\label{eq:eoms_var_spectators}
E_{\mu a }l^a_i \bar{\partial} X^i& - E_{a \mu} l^a_i \partial X^i + E_{\mu \nu} \bar{\partial} Y^\nu - E_{\nu \mu} \partial Y^\nu \nonumber \\
&=\tilde{E}_\mu^{\ a} \tilde{l}_{ai} \bar{\partial} \tilde{X}^i - \tilde{E}^a_{\ \mu} \tilde{l}_{ai} \partial X^i + \tilde{E}_{\mu \nu} \bar{\partial} Y^\nu - \tilde{E}_{\nu \mu} \partial Y^\nu\,.
\end{align}
The first two equations can be solved by making an Ansatz similar to that of the non-spectator case  but taking now into account a contribution from the spectator fields. For example the first line in the particular solution given in eq. \eqref{eq:Eab on defect} now becomes
\begin{align}\label{eq:Eab on defect_spectators}
E_{a\mu} \bar{\partial} Y^\mu + \left(E_{ab} + C_a^{\ c}\tilde{\Pi}_{cb}[\tilde{g}] \right) l^b_j \bar{\partial} X^j &=  -   C_a^{\ b} \tilde{l}_{bj}  \bar{\partial}  \tilde{X}^j\,,
\end{align}
and mutatis mutandis for the three other equations. Again, by requiring consistency of these particular solutions leads to the relation  
\begin{align*}
\tilde{E}^{ab} + \tilde{C}^a_{\ c}\Pi^{cb}[g] \equiv  \tilde{C}^a_{\ d} \left( (E + C\tilde{\Pi}[\tilde{g}] )^{-1} \right)^{bc} C_c^{\ d}\,, \quad \tilde{E}^a_{\ \mu}  \equiv  \tilde{C}^a_{\ d} \left( (E + C\tilde{\Pi}[\tilde{g}] )^{-1} \right)^{bc} E_{c\mu}\,,
\end{align*}
and similarly for the others. We are again lead to conclude that the defect implements the canonical transformation for the Poisson-Lie T-duality, but this time including spectator directions.

It remains to show that the third equation of motion coming from the spectators in eq. \eqref{eq:eoms_var_spectators} is also satisfied for the specific solution singled out by the Ansatz. To this purpose we have to use the form of the background field imposed by the Poisson-Lie symmetry of the model as given in eqs \eqref{eq:gen_E_PL_spect}. With these expressions at hand, the canonical transformation relating the left-invariant forms of the Poisson-Lie T-duality  backgrounds on $M$ and $\tilde M$ can be written as  \cite{Sfetsos:1996xj}
\begin{align}\label{eq:canonical_transfo_spect}
(E_0^\mp)^{-1} E^\mp (L_\pm \pm \Pi F^\mp \partial_\pm Y) = \pm   \tilde{E}^\mp (\tilde{L}_\pm \mp F^\mp \partial_\pm Y)\,.
\end{align}
A little algebra shows that this is equivalent to the relation 
\begin{align}\label{eq:canonical_transfo_spect_2}
 E_{a \mu}l^a_i \partial X^i  &=   \tilde{E}^a_{\ \mu} \tilde{l}_{ai} \partial \tilde{X}^i-  F_{\nu a} \tilde{E}^a_{\ \mu}\partial Y^\nu  +  F_{\nu a} \Pi^{ab} E_{a \mu}\partial Y^\nu\,.
\end{align}
Using the rewriting of the canonical transformation for Poisson-Lie T-duality in eq. \eqref{eq:canonical_transfo_spect_2} together with the general expressions for the background fields in eq. \eqref{eq:gen_E_PL_spect} imposed by Poisson-Lie symmetry, one can then easily show that the third equation of motion coming from the variation of the spectators in eq. \eqref{eq:eoms_var_spectators} is also satisfied. Note finally that in this case the pair $(\mathcal Y,\mathcal F)=(\mathbbsl D,\omega_\mathrm{STS})$ no longer defines a space-filling defect since the corresponding worldvolume is again the Drinfel'd double $\mathbbsl D\subset \mathcal G\times \tilde{\mathcal G}\times N$.

Lastly, we turn to the problem of showing that this defect is topological. Taking into account the spectator directions, the energy momentum tensors featured in the topological defect condition in eq. \eqref{eq:top_def_cond} can be written as (wlog, for the action on the left)
\begin{align*}
T = G_{ab}l^a_i l^b_j \partial X^i \partial X^j + 2 G_{a\mu} l^a_i \partial X^i \partial Y^\mu + G_{\mu \nu} \partial Y^\mu \partial Y^\nu\,.
\end{align*}
We will now show that demanding again that the curvature of the one-form field $\mathbbsl A$ living on the defect line is the Semenov-Tian-Shansky symplectic form, guarantees the cancelation of the holomorphic and anti-holomorphic parts of the energy-momentum tensors. Using the expressions for the background fields in eq. \eqref{eq:gen_E_PL_spect}, the difference of the holomorphic parts of the energy momentum tensors on both sides reads 
\begin{align*}
T-\tilde{T} &= E_{a\mu}  l^a_i \partial X^i \partial Y^\mu - \tilde{E}^a_{\ \mu} \tilde{l}_{ai} \partial \tilde{X}^i \partial Y^\mu - F_{\mu a }\Pi^{ab}E_{b \nu} \partial Y^\mu \partial Y^\nu  + F_{\mu a }\tilde{E}^a_{\ \nu} \partial Y^\mu \partial Y^\nu\,,
\end{align*}
where the antisymmetry of the combinations $C\Pi$ and $\tilde C\tilde \Pi$ was used to cancel a number of terms contributing to the left- and right-energy momentum tensors.
But this last line however is nothing but the canonical transformation for Poisson-Lie T-duality in presence of spectator directions as derived in eq. \eqref{eq:canonical_transfo_spect_2}. An identical computation shows that the anti-holomorphic components of energy momentum tensors of the theories on the left and the right of the defect cancel as well, concluding that the defect does not only encode Poisson-Lie T-duality with spectators but is in fact topological. In what follows, we will often refer to this topological defect $(\mathbbsl D,\mathcal F)$ encoding Poisson-Lie T-duality by the shorter nomicker ``Poisson-Lie defects''.

%%%%%%%%%%%%%%%%%%%%%%%%%%%%%%%%%%%%%%%%%%%	
\subsubsection{Relation to Fourier-Mukai-like transforms}\label{sec:FM}
One of our motivations to study whether or not Poisson-Lie T-duality could also be realised by a topological defect was to better understand if this generalised form of T-duality can be associated to a Fourier-Mukai-like transform. It has been known since \cite{Hori:1999me} that the T-duality transformation of the Ramond-Ramond fields is encoded through a Fourier-Mukai transform. The latter is an integral transform defining an isomorphism of, possibly $H$-twisted, cohomologies for the manifolds $M_1$ and $M_2$ and takes the formal form
\begin{align}\label{eq:FM_abstr}
	\Lambda^\bullet T^\ast M_1\rightarrow \Lambda^\bullet T^\ast M_2: \alpha \mapsto (p_2)_\ast(K\cdot p_1^{\ast} \alpha )\,,
\end{align}
here $p_i:M_1\times M_2\rightarrow M_i$ is the projection map onto $M_i$ and $K$ is called the kernel of the integral transform. That is, differential forms on the manifold $M_1$ are first transported to the product space $M_1\times M_2$ by the pull-back of the projection map $p_1:M_1\times M_2\rightarrow M_1$. In the product space $M_1\times M_2$ the pull-backed object is then ``convoluted'' with a specific kernel $K$ encoding the duality transformation or isomorphism. The result is then pushed forward via the projection $p_2:M_1\times M_2\rightarrow M_2$ onto the second factor, to a ``dual'' differential form defined on the manifold $M_2$. Schematically:
 \tikzset{every picture/.style={line width=0.6pt}} 
\begin{equation}    
\begin{aligned}
\begin{tikzpicture}[x=0.75pt,y=0.75pt,yscale=-1,xscale=1]
\draw    (294.14,46.39) -- (276.16,76.67) ;
\draw [shift={(275.14,78.39)}, rotate = 300.7] [color={rgb, 255:red, 0; green, 0; blue, 0 }  ][line width=0.75]    (10.93,-4.9) .. controls (6.95,-2.3) and (3.31,-0.67) .. (0,0) .. controls (3.31,0.67) and (6.95,2.3) .. (10.93,4.9)   ;
\draw    (320.14,47.39) -- (341,77.74) ;
\draw [shift={(342.14,79.39)}, rotate = 235.49] [color={rgb, 255:red, 0; green, 0; blue, 0 }  ][line width=0.75]    (10.93,-4.9) .. controls (6.95,-2.3) and (3.31,-0.67) .. (0,0) .. controls (3.31,0.67) and (6.95,2.3) .. (10.93,4.9)   ;

% Text Node
\draw (277,27.24) node [anchor=north west][inner sep=0.75pt]    {$M_1\times M_2$};
% Text Node
\draw (258,83.24) node [anchor=north west][inner sep=0.75pt]    {$M_1$};
% Text Node
\draw (337.14,83.24) node [anchor=north west][inner sep=0.75pt]    {$ M_2$};
% Text Node
\draw (260,48.24) node [anchor=north west][inner sep=0.75pt]    {$p_{1}$};
% Text Node
\draw (337,48.24) node [anchor=north west][inner sep=0.75pt]    {$p _{2}$};
\end{tikzpicture}
\end{aligned}
\end{equation}
The question to whether generalisations of T-duality might also admit a representation as a Fourier-Mukai transform is a challenging task. Foremost, since Fourier-Mukai transformations between spaces that are not associated to abelian varieties\footnote{ In fact, a Fourier-Mukai transform is an equivalence of derived category of coherent sheaves. This is realised in topological string theory, where B-branes are modelled by coherent sheaves and the relevant kernel to encode Abelian T-duality is given by the curvature of the Poincar\'e bundle. Whether such a neat picture could also arise for the non-Abelian or Poisson-Lie generalisations of T-duality is at present unclear to the authors. } are not well-understood.  Remarkably, topological defects associated to dualities offer a way around this technical and conceptual hurdle.

 In \cite{Bachas:2001vj,Bachas:2012bj,Kapustin:2009av,Sarkissian:2008dq}, it was noticed that the Fourier-Mukai transform naturally surfaces when studying the worldsheet topological defects encoding Abelian T-duality. Later, using the target space formulation of topological defects, the kernels of the Fourier-Mukai transforms for non-Abelian \cite{Gevorgyan:2013xka} and fermionic T-duality \cite{Elitzur:2013ut} were shown to again be essentially captured by the data defining the duality defect. This relation between topological defects and a Fourier-Mukai like operation is not coincidental and is a direct consequence of the existence of a fusion operation of T-duality inducing topological defects.  Applying a T-duality to a D-brane should intuitively equate to fusing the pertinent defect together with the boundary condition. As such, the objects defining the duality defect have to also contain the necessary information to understand how the Ramond-Ramond fields, which are charged under the D-brane, transform under T-duality. More precisely, the authors in \cite{Sarkissian:2008dq,Gevorgyan:2013xka} showed how the two-form $ \mathcal F$ living on the worldvolume of the topological defect encoding Abelian (and later for non-Abelian) T-duality determines the kernel $K$ of a Fourier-Mukai-like integral transform acting on the Ramond-Ramond fields via  $\int_{M_1}  \exp( F_\mathrm{kernel}\cdot p_1^{\ast} \alpha )$ where $ F_\mathrm{kernel}=-(p^\ast B-\tilde p^\ast \tilde B+\mathcal F)$, where we recall that $\mathcal F$ is the curvature of the line bundle on the topological defect encoding the duality.

 In addition, in \cite{Bachas:2012bj}, see also \cite{Kapustin:2009av}, the authors argued that fusion turns out to match the composition of the associated geometric integral transformations in all studied examples. The authors also argue that the fusion of a defect with two-form curvature $\mathcal F$ with any D-brane $\mathcal Y_1$  determines a transformation that maps D-branes $\mathcal Y_1$ in $M_1$ to D-branes given by
\begin{align}\label{eq:str_fusion_FM}
	(p_2)_\ast (\mathcal F\cdot p_1^\ast \mathcal Y_1)
\end{align} 
on $M_2$. This observation, which was first suggested in \cite{Bachas:2001vj} and later studied in e.g. \cite{Bachas:2012bj,Fuchs:2007fw}, will return later when we consider the notion of fusion.
 
The discussion in the previous sections suggests, following the same logic, that one should identify the kernel of the integral transform associated to Poisson-Lie T-duality with the Semenov-Tian-Shansky symplectic structure of the double $\mathbbsl{D}$. This observation nicely supports\footnote{Note however that, at least naively, this form for the kernel of the Fourier-Mukai transform encoding Poisson-Lie T-duality differs from the one proposed in \cite{Demulder:2018lmj}.} the results of \cite{Arvanitakis:2021lwo}, where, by studying the global aspects of Poisson-Lie T-duality, the Semenov-Tian-Shansky symplectic structure was also proposed as the relevant kernel of the Fourier-Mukai transform for Poisson-Lie T-duality. In \cite{Arvanitakis:2021lwo}, an expression for the kernel of Poisson-Lie T-duality \textit{with} spectator fields was also derived by studying global aspects of the generalised T-duality and imposing the Bianchi identity on the result of the associated Fourier-Mukai transformation. This new kernel describing the duality with spectators displays additional terms that account for the non-trivial fibration of spectator direction across the manifold. This information can only be detected when considering the global and topological aspects of Poisson-Lie T-duality. In the present context however, since the topological defect encodes a local version of the generalised T-duality transformations, these terms remain invisible. This explains why, for our purposes, the additional terms do not play a role in the construction of the associated defect. 

%%%%%%%%%%%%%%%%%%%%%%%%%%%%%%%%%%%%%%%%%%%
%%%%%%%%%%%%%%%%%%%%%%%%%%%%%%%%%%%%%%%%%%%
\subsection{Product space and generalised geometry approach }
As was already mentioned in the introduction, topological defects have the distinguishing feature that they can be deformed continuously from their initial location. When this defect line is superimposed with another defect line (or a boundary condition), they ``fuse'' and one obtains a new configuration: either a new defect (or new boundary condition) or a superposition of multiple defects (or boundary conditions). How worldsheet fusion translate at the level of the target space will be subject of section \ref{sec:fusion}. To prepare the ground to understand how to realise ``target space fusion'', i.e. an operation on the set of submanifolds with two-form field $(\mathcal Y,\mathcal F)$ on the product space $M\times \tilde M$ that defines topological defects, it turns out to be instrumental to reformulate the notion of bibranes in the setting of generalised geometry. 

Indeed, topological defects, using the language of generalised geometry, can be equally characterised directly in terms of the background fields and the pair $(\mathcal Y,\mathcal F)$ capturing the relevant defect \cite{Kapustin:2010zc}. We will first review here some of their results useful for our purposes. As was already stressed in the previous section, on the target space the defect is described through the corresponding submanifolds in the doubled space $M\times \tilde M$ which carries a two-form $\mathcal F$. The condition for the defect to be topological, see eqs. \eqref{eq:top_def_cond}, can however be directly expressed in terms of conditions on the submanifold $\mathcal Y$ and its two-form $\mathcal F$. To do so, one first introduces a neutral signature metric $\mathbbsl{G}_{AB}$ and B-field $\mathbbsl{B}_{AB}$ on the product space $M\times \tilde M$ given by
\begin{align*}
\mathbbsl{G}_{AB} = \begin{pmatrix}
G_{ab} & 0\\
0 & - \tilde{G}_{ab}
\end{pmatrix}\,, \qquad \mathbbsl{B}_{AB}=
\begin{pmatrix}
B_{ab} & 0\\
0 & -\tilde{B}_{ab}
\end{pmatrix}\,,
\end{align*}
where $(G,B)$, respectively $(\tilde G,\tilde B)$, are the background fields of the sigma-models on the left, respectively right, of the defect, implicitly restricted to the boundary $\mathcal Y$ where they are well-defined. The capital index $A$ runs form $1$ to $2n$ with the first $n$ indices belonging to the target space $\mathcal G$ and the latter $n$ to the space $\tilde{\mathcal G}$, for $n=\dim \mathcal G=\dim \tilde{\mathcal G}$. Furthermore we define a two-form $\mathbbsl{F}$ to be the combination
\begin{align}\label{eq:calF}
\mathbbsl{F}  = - \mathrm d \mathbbsl{A} - \iota^\ast \mathbbsl{B}= -\mathcal F - \iota^\ast \mathbbsl{B}
\end{align}
with $\iota: \mathcal Y \hookrightarrow \mathcal G \times \tilde{\mathcal G}$  is the embedding map and $\mathcal F$ is the field strength of the one-form $\mathbbsl A$ living on the defect worldvolume $\mathcal Y$. Kapustin and Setter showed in \cite{Kapustin:2010zc} that, the condition on the energy momentum tensor can be translated to a condition on the geometry of the induced target space worldvolume together with the background fields. In order to define allowed pairs $(\mathcal Y,\mathcal{F})$ one has to verify the relation\footnote{In \cite{Kapustin:2010zc}, pairs determining  defects are chosen to be  characterised by the curvature \textit{plus} the B-field contribution. Note that in the present paper, we have picked a slightly different convention having chosen to characterise the pair by $(\mathcal Y,\mathcal F)$ where $\mathcal F$ is the curvature of the line bundle, without the B-fields. }
\begin{align}\label{eq:top_def_cond_product_sp}
\left(\left. \mathbbsl{G} \right|_{S\mathcal Y}^{-1}  \left. \mathbbsl{F}\right|_{S\mathcal Y}\right)^2 = \mathds{1}\quad \text{and}\quad  \mathrm{ker}\,\mathbbsl F=(T\mathcal Y)^\perp\,,
\end{align}
where $S\mathcal Y$ is the so-called screen distribution since we are working in a non-Riemannian signature\footnote{See \cite{Kapustin:2010zc,bejancu1995lightlike}.
Since we work in a pseudo-Riemannian manifold, one can no longer simply split the generalised tangent space in the usual way $\iota^\ast TM = T\mathcal Y \oplus (T\mathcal Y)^\perp$.  
When $\mathcal Y$ is a coisotropic submanifold, i.e. $T\mathcal Y\supset (T\mathcal Y)^\perp$, however, one can show that there still exists a canonical spitting, albeit by introducing a new object. Indeed, in this case the tangent space has to be  complemented by the so-called screen distribution $S(T\mathcal Y)$, $T\mathcal Y=S(T\mathcal Y)\oplus (T\mathcal Y)^\perp$, where here the direct sum is between transverse spaces. The (pull-back) of the manifold $\mathbbsl M$ is then decomposed as \cite{bejancu1995lightlike}
\begin{align}
	\iota^\ast T\mathbbsl M= T\mathcal Y\oplus N\mathcal Y= S(T\mathcal Y)\oplus (T\mathcal Y)^\perp\oplus N\mathcal Y\,,
\end{align}
where $N\mathcal Y$ is a complementary transverse distribution to $T\mathcal Y$ in $\iota^\ast TM$.}. This equation provides an equivalent way to characterise topological defects at the level of the product space $M\times \tilde M$ purely in terms of background and target space data. In what follows we will refer to the topological defect condition given  as in eq. \eqref{eq:top_def_cond_product_sp}, as the ``target space condition''.
When the submanifold $\mathcal Y$ corresponding to the defect is space-filling, i.e $\mathcal Y=\mathcal G \times \tilde{\mathcal G}$ we have $(T\mathcal Y)^\perp = 0$ and the condition simplifies to $\ker \mathbbsl{F} =0$ and $(\mathbbsl{G}^{-1}\mathbbsl{F})^2=\mathds{1}$. Note that in this case, this also implies that $\mathbbsl F$ is a symplectic structure on the product space $\mathcal G\times \tilde{\mathcal G}$. Before reviewing how this condition can be expressed in generalised geometry, let us show again that the space-filling defect with two-form field given by the Semenov-Tian-Shansky symplectic form induces Poisson-Lie T-duality and is topological, however using this time the target space condition above.

%%%%%%%%%%%%%%%%%%%%%%%%%%%%%%%%%%%%%%%%%%%
\subsubsection{Poisson-Lie T-duality topological defect -- revisited}\label{sec:top_def_product_space}
Having introduced the target space condition, given in eq. \eqref{eq:top_def_cond_product_sp}, which warrants that a defect pair $(\mathcal Y,\mathcal F)$ is topological, we give an alternative way to show that the pair $(\mathcal Y,\mathcal F)=(\mathbbsl D,\omega_\mathrm{STS})$ defines in fact a topological defect inducing Poisson-Lie T-duality.  From the Semenov-Tian-Shansky symplectic structure, using the expression given in eq. \eqref{eq:STS_worked_out}, we can read off the matrix representation of $\mathrm d \mathbbsl A$. The associated two-form field $\mathbbsl{F}$ is then given, as in section \ref{sec:sigma_model_approach}, by 
\begin{align}
\mathbbsl{F} =  -\mathrm d\mathbbsl{A} - \iota^\ast \mathbbsl{B} = \begin{pmatrix}
-C \tilde{\Pi} - B & -C\\
\tilde{C} &  \tilde{C} \Pi + \tilde{B}
\end{pmatrix}\,.
\end{align}
We can now directly compute $(\mathbbsl{G}^{-1}\mathbbsl{F})^2$ using the definition for $\mathbbsl F$ in eq. \eqref{eq:calF}, yielding the (1,1)- and (1,2)-block components respectively
\begin{align*}
[(\mathbbsl{G}^{-1}\mathbbsl{F})^2]_{11}&=\left(G^{-1}(C \tilde{\Pi} +B)\right)^2 + G^{-1}C\tilde{G}^{-1}\tilde{C}\,, \\  [(\mathbbsl{G}^{-1}\mathbbsl{F})^2]_{12}&=G^{-1}(C \tilde{\Pi} +B)G^{-1}C +  G^{-1}C \tilde{G}^{-1}(\tilde{C} \Pi+\tilde{B})\,.
\end{align*}
It is instructive to first examine the simpler case of Abelian T-duality. Specialising to Abelian T-duality means that we have that $\Pi=0, \tilde{\Pi}=0$ and the transformation rules relating the background fields and their T-dual are known to take on the form \cite{Giveon:1994fu}
\begin{align}
\tilde{G}=(G-BG^{-1}G)\,, \quad  \tilde{G}^{-1}\tilde{B} = - B G^{-1}\,.
\end{align}
One can readily see that in this case we indeed have that $(\mathbbsl{G}^{-1}\mathbbsl{F})^2 =\mathds 1$, and the defect describes Abelian T-duality and is in addition topological. In a very similar way, using the eqs. \eqref{eq:dual_bckgrd_ito_original} derived  from the Poisson-Lie canonical transformation  in the Appendix \ref{app:Canonical_PL} relating the Poisson-Lie background fields $E=G+B$ and its dual background field $\tilde E=\tilde G+\tilde B$, one can verify after a little algebra that the defect $(\mathbbsl D, \omega_\mathrm{STS})$ that captures now generalised T-duality is again topological, consistent with what was established already in section \ref{sec:Lagr_def_no_spect} at the level of the action. 

%%%%%%%%%%%%%%%%%%%%%%%%%%%%%%%%%%%%%%%%%%%
\subsubsection{Topological defects as Dirac structures}\label{sec:top_def_GG}
We now turn to the generalised geometry formulation of topological defects for non-linear sigma-models as was originally given in \cite{Kapustin:2010zc}. Before reviewing the generalised geometry condition for topological defect, we would like to anticipate on our discussion about fusion in section \ref{sec:fusion} and introduce the notion of Dirac structure in generalised geometry. To do so, we will explain how D-branes and the worldvolume of topological defects can be  described using the same objects. This can be achieved precisely by promoting their description to generalised geometry, or rather Dirac geometry. As we will shortly review, Dirac geometry unifies symplectic structures, Poisson structures and foliations in the framework of generalised geometry.

The crucial observation is to note that the worldvolume of topological defects and D-branes are both determined not only by a submanifold but have to be supplemented by a line-bundle with curvature $\mathcal F$ on its worldvolume. As we will now review, this information is exactly encoded by Dirac structures. A Dirac structure is a Lagrangian subspace $L$ of a Courant algebroid\footnote{We refer the reader to appendix \ref{app:GG} for definitions and a short introduction to topics in generalised geometry relevant to this section.} $E$ over a manifold $M$ that is involutive with respect to the Courant bracket on $E$. The latter condition guarantees that the projection of the Dirac structure onto the tangent space at each point of the manifold $M$ defines an integrable distribution. That is, the Dirac structure $L$ on $M$ defines a stratification of $M=\sum_i \mathcal Y_i$, where each leaf $\mathcal Y_i$ inherit a two-form field $\mathcal F_i$, see e.g. \cite{gualtieri2011generalized}. 

A basic example is well-known. A Poisson-structure $\Pi$ on a manifold $M$ determines a foliation of $M$ into symplectic leaves on which the Poisson-structure $\Pi$ is invertible. Each such leaf $\mathcal Y$ of the symplectic foliation thus carries a two-form field $\omega=(\pi|_{\mathcal Y})^{-1}$; the inverse of the Poisson structure restricted to that leaf. When the Poisson structure is non-degenerate on $M$, the foliation consists of just one leaf, the whole manifold $M$ and it is equivalent to a symplectic structure. The Dirac structure in these examples coincide with the graph of Poisson structure or the symplectic structure in the generalised tangent space, see appendix \ref{app:GG}. A general Dirac structure $L$ can in fact always be identified with a subbundle that corresponds to the graph of a two-form $ \mathcal F$ whose projection to the tangent bundle is an integrable distribution for a submanifold $\mathcal Y$. That is we have that any Dirac structure $L$ can be identified with a subbundle
\begin{align}\label{eq:DS_gen_form}
	L(\mathcal Y, \mathcal F)= \left\{ (X, \xi) : X \in T\mathcal Y, \xi \in T^\ast M  \ \text{such that} \ {\mathcal F}X=\iota^\ast \xi \right\}\,,
\end{align}
where $\iota$ again denotes the inclusion map of $\mathcal Y$ into $M$. Conversely, a couple $(\mathcal Y,\mathcal F)$ of a submanifold $\mathcal Y\subset M$ and a  two-form $\mathcal F$ can be shown to define a Dirac structure $L$ of $E=\mathbb TM$. 
 
We can now give some notable examples of Dirac structure describing D-branes. In WZW models, (twisted) D-branes  correspond to so-called Cartan-Dirac structure carrying the GKHJW-structure \cite{guruprasad1997group} and whose leafs are conjugacy classes. More generally, D-branes described by a particular subset of Dirac structures have also been studied in \cite{Asakawa:2012px,Asakawa:2014eva}. In \cite{klimvcik1996poisson}, D-branes in Poisson-Lie models were identified with the symplectic leaves of the Poisson structure associated to the Lie group $M=\mathcal G$. In \cite{Severa:2015hta}, by realising that variational problems associated to two-dimensional sigma-models are associated to exact (equivariant) Courant algebroids, boundary conditions translate into the choice of a Dirac structure $L$ in that Courant algebroid $E$ over $M$. 

For topological defects, it was shown in \cite{Kapustin:2010zc}, although not explicitly coined as such, that topological defects can be described as Dirac structures $L$ now living in a Courant algebroid $E$ over the product space $M\times \tilde M$, satisfying an additional compatibility condition. Let us review the argument. Define a new ``generalised metric'' $\mathcal{R}_{\mathbbsl G}$ for the product space $\mathbbsl M=M\times \tilde M$ given by
\begin{align}
\mathcal{R}_\mathbbsl{G}  = \begin{pmatrix}
0 & \mathbbsl{G} ^{-1}\\
\mathbbsl{G}  &0
\end{pmatrix}\,,
\end{align}
defining a generalised almost product structure, i.e an endomorphism $\mathcal{R}_{\mathbbsl G}:T\mathbbsl{M} \oplus T^\ast \mathbbsl{M} \to T\mathbbsl{M} \oplus T^\ast \mathbbsl{M}$ such that $\mathcal{R}^2_{\mathbbsl G}=\mathrm{id}$.
As we remarked earlier, the choice of a foliation together with the associated two-form is equivalent to having a Dirac structure of the form given in eq. \eqref{eq:DS_gen_form}.
Then the pair $(\mathcal Y,\mathcal{F})$, consisting of a  two-form $\mathcal{F}$ on $\mathcal Y \subseteq M\times \tilde M$ a submanifold, is a topological defect if and only if\footnote{Its not hard to see that this equivalence: First let $(\mathcal Y,\mathcal{F})$ such that $\mathcal{R}_\mathbbsl{G} (L(\mathcal Y,\mathcal F)) = L(\mathcal Y,\mathcal F)$ and let $(X,\xi) \in L(\mathcal Y,\mathcal F)$ arbitrary. Then $\mathcal{F}X=\iota^\ast \xi$ and there exists some $Z \in \iota^\ast T\mathbb{M}$ such that $\iota^\ast \xi = \mathbbsl{G} Z$. Therefore $\mathcal{R}_\mathbbsl{G} (X, \xi) = (\mathbbsl{G}^{-1}\mathbbsl{G}Z, \mathbbsl{G}) = (Z,\mathbbsl{G}X) \in L(\mathcal Y,\mathcal F)$, which implies $\mathcal{F}Z=\mathbbsl{G}X$ and $Z \in T\mathcal Y$ and therefore $(\mathcal Y,\mathcal{F})$ is a topological bibrane. The converse direction requires bit more work, see \cite{Kapustin:2010zc} for details.} the following holds 
\begin{equation*}
\mathcal{R}_{\mathbbsl G} (L(\mathcal Y,\mathcal F)) = L(\mathcal Y,\mathcal F) \,.
\end{equation*}
That is topological defects correspond to Dirac structures of a Courant algebroid over $M\times \tilde M$ that are stable under the product structure $\mathcal R_{\mathbbsl G}$. This formulation is indeed reminiscent of how D-brane compatible with extended supersymmetry \cite{Zabzine:2004dp}, D-branes in para-Hermitian geometry \cite{Marotta:2022tfe} or complex submanifolds in generalised complex geometry \cite{gualtieri2011generalized} are characterised.

%%%%%%%%%%%%%%%%%%%%%%%%%%%%%%%%%%%%%%%%%%%
%%%%%%%%%%%%%%%%%%%%%%%%%%%%%%%%%%%%%%%%%%%
%%%%%%%%%%%%%%%%%%%%%%%%%%%%%%%%%%%%%%%%%%%
\section{Worldvolume fusion of generalised T-duality defects}	\label{sec:fusion}
In this section we turn to the question of how the fusion of topological defects can be understood as an operation on the target space. In contrast to worldsheet fusion,  very little is known on how to consistently define fusion in target space. First insights came from studying the particular case of  fusing topological defects (or also called bibranes) in WZW models \cite{Fuchs:2007fw}. As can be expected from the study of worldsheet fusion, target space fusion of two topological defects or bibranes $(\mathcal Y_i,\mathcal F_i)_{i=1,2}$ results in single or possibly a superposition of multiple defects,  schematically 
	\begin{align}\label{eq:fusion_sum}
		(\mathcal Y_1,\mathcal F_1)\star(\mathcal Y_2,\mathcal F_2)=\sum_\alpha N_{12}{}^\alpha (\mathcal Y_\alpha,\mathcal F_\alpha)\,,
	\end{align} 
where $ N_{12}{}^\alpha$ are some constant coefficients for a collection of bibranes $(\mathcal Y_\alpha,\mathcal F_\alpha)$. The transparent or invisible defect, which separates $M$ with itself, acts as the unit element of the fusion operation. Then a defect is invertible for the operation of fusion when it can be fused with a second to yield the invisible defect. Remarkably, it turns out that for WZW models the fusion operation was suggested to yield the  Verlinde algebra \cite{Fuchs:2007fw}. In general however, the fusion operation $\star$ will not define a group operation as some elements might fail to have an inverse or the algebra might not even close.

In what follows, we first set  out to understand the fusion properties of the Poisson-Lie T-duality defect derived earlier. We  provide a fusion realisation of the generalised T-duality transformations of boundary conditions characterised in terms of a gluing matrix. Subsequently, we will address the effect of fusion at the level of the target space. Combining a Fourier-Mukai like operation and the observation that Dirac structures both encode the worldvolume data of defects and D-brane, we write down a target space fusion operation encoding the D-brane resulting from the fusion of defect and D-brane.

%%%%%%%%%%%%%%%%%%%%%%%%%%%%%%%%%%%%%%%%%%%
%%%%%%%%%%%%%%%%%%%%%%%%%%%%%%%%%%%%%%%%%%%
\subsection{Lagrangian fusion and gluing conditions}\label{sec:BC_gluing}
In this section, we first examine the result of fusing a generalised T-duality defect with a boundary condition at the Lagrangian level. We first remind the reader of how boundary conditions can be encoded using a gluing matrix. We then show how fusing the Poisson-Lie topological defect in this way correctly predicts the transformation rules under Poisson-Lie T-duality for the boundary conditions of open strings  in terms of a gluing matrix. 
 
%%%%%%%%%%%%%%%%%%%%%%%%%%%%%%%%%%%%%%%%%%%
\subsubsection{Gluing matrices and Dirac structures}
A convenient way to encode boundary conditions for open strings is by using so-called gluing matrices \cite{Albertsson:2001dv}. The boundary condition for open strings are then imposed by demanding that the left- and right-moving worldsheet fields are related on the boundary by a gluing operator $ R$:
\begin{align}\label{eq:gluing_conds}
	\partial_-X|_{\sigma=0,\pi}=R|_{\sigma=0,\pi}\cdot \partial_+X|_{\sigma=0,\pi}\,.
\end{align} 
The matrix $R$ cannot be arbitrary and has to satisfy a number of conditions reflecting the consistency of the underlying boundary conditions. In particular, conformal invariance requires the matrix $R$ to preserve the metric. The Dirichlet directions, those normal to the D-brane, correspond to the $-1$ eigenspace of the matrix $R$. These can be singled out by the projector $RP_\parallel=P_\parallel R=-P_\parallel$. In addition, to warrant for the existence of a smooth underlying submanifold modelling the D-brane on the target space, the direction along the orthogonal directions obtained from the projector $P_\perp=\mathds 1-P_\parallel$, encoding the Neumann directions, should generate an integrable distribution and a corresponding foliation of the manifold $M$. On each such leaf $\mathcal Y\subset M$ of the foliation, there exists a two-form $\mathcal F$ such that $\mathrm d \mathcal F=H|_{\mathcal Y}$.  

Under a T-duality transformation, the gluing matrix $R$ capturing the specific boundary conditions is mapped to a new, dual, gluing matrix $\tilde R$ for the boundary conditions in the dual space. How the gluing matrix transforms under the action of Poisson-Lie T-duality was derived in \cite{Albertsson:2006zg,Albertsson:2007it}. We will show in this section that the fusion operation of the Poisson-Lie T-duality defect $(\mathbbsl D,\omega_\mathrm{STS})$ with a boundary condition encoded by a gluing matrix $R$ consistently yields the expected Poisson-Lie T-dual gluing matrix.

For later use, let us however first review how boundary conditions arising from this type of gluing conditions can be formulated within generalised geometry. As pointed out in \cite{Zabzine:2004dp}, the gluing condition can be reformulated in an $O(d,d)$ covariant way. Introduce a ``doubled'' gluing matrix on the generalised tangent space $\mathcal R:\mathbb TM\rightarrow \mathbb TM$ which is taken to be an $O(d,d)$ element. This $O(d,d)$-valued gluing matrix $\mathcal R$ is defined in terms of the projectors induced by the gluing matrix $R$ acting on the tangent space via
\begin{align*}
	\mathcal R=\begin{pmatrix}
		P_\Delta & 0\\ \mathcal FP_\Delta+P_\Delta^T\mathcal F & -P_\Delta^T
	\end{pmatrix}\,,
\end{align*}
where $P_\Delta =(P_{\parallel}-P_\perp)$ and the two-form $\mathrm d \mathcal F=H|_{\mathcal Y}$ on a leaf $\mathcal Y$ of the foliation. This gluing matrix is unipotent $\mathcal R^2=1$ and has to be compatible with the generalised metric.
From the consistency condition of the conventional gluing matrix $R$ above, one should have that the $+1$-eigenspace generate the integrable distribution of the worldvolume of the D-brane when restricted to the tangent space. In particular, the projector $\tfrac{1}{2}(1-\mathcal R)$ can be used to define a Dirac structure $L(\mathcal Y,\mathcal F)\subset  \mathbbsl TM$ given by the expression 
\begin{align}\label{eq:Dirac_str_gluing_conds}
	L(\mathcal Y,\mathcal  F)=\{\tfrac{1}{2}(1-\mathcal R)(X+\xi)=(X+\xi)\mid X+\xi\in T\mathcal D \oplus T^\ast M|_\mathcal{D}\,,\; \xi|_\mathcal{D}=\iota_X\mathcal  F \}\,.
\end{align}
That is, the boundary condition characterised by the gluing matrix in eq. \eqref{eq:gluing_conds} can be, on the target space, equivalently be described in terms of the Dirac structure in eq. \eqref{eq:Dirac_str_gluing_conds}.

%%%%%%%%%%%%%%%%%%%%%%%%%%%%%%%%%%%%%%%%%%%
\subsubsection{Boundaries under Poisson-Lie T-duality via fusion}
We will now show how to obtain the Poisson-Lie T-dual boundary conditions by fusing boundary condition of the form given in eq. \eqref{eq:gluing_conds} with the topological defect encoding Poisson-Lie T-duality that was put forward in the first section.
The set-up is the same as in section \ref{sec:Lagr_def_no_spect}, with the additional assumption that the system on, say the left, is subject to boundary condition described by the gluing condition as in eq. \eqref{eq:gluing_conds} for a given gluing matrix $R$\footnote{Note that this $R$ should not be confused with the right-invariant Maurer-Cartan of $\mathcal G$. Since in this section, we will solely work with the left-invariant form $L=l_idX^i$, there should be no source of confusion.}. Since the Poisson-Lie T-duality defect is topological we can (formally) move the defect line from its initial position at $\sigma=0$ on the worldsheet to the boundary of the left system. At the boundary we now both have to impose the boundary conditions and the defect equations. Remembering the equations of motion taking into account the defect contribution given in eq. \eqref{eq:EOMs_at_defect},
these relations however have to be supplemented with the boundary conditions encoded by the gluing matrix, i.e. $L_-=RL_+$. Plugging this gluing condition into the first equation of motion in eq. \eqref{eq:EOMs_at_defect}, leads to 
\begin{align*}
\tilde{l}_{bj} \partial_\tau \tilde{X}^j = (C^{-1})_{b}^{\ a} E^R_{ac}l^c_j \partial X^j - \tilde{\Pi}_{bc}[\tilde{g}] l^c_j \partial_\tau X^j\,,
\end{align*}
where we have defined the combination $E^R\equiv E^T   - E R$. Substituting this last relation into the equation of motion of the dual model in eq. \eqref{eq:EOMs_at_defect} and taking into account the relation  $\Pi C^{-1}= \tilde{C}^{-1}\tilde \Pi$, leads to 
\begin{align}
l^a_j \partial_\tau X^j &= \tilde{E}^{b a} \tilde{l}_{b j} \partial \tilde{X}^j  - \tilde{E}^{a b} \tilde{l}_{b j} \bar{\partial}\tilde{X}^j - \Pi^{ab}  E^R_{bc} l^c_j \partial X^j\,.
\end{align}
Using the expression $l^a_j \partial_\tau X^j=L_++L_-$ and enforcing that the Maurer-Cartan forms on $\mathcal G$ and $\tilde{ \mathcal G}$ are related by the canonical transformation for Poisson-Lie T-duality given in eq. \eqref{eq:canonical_transfo_Ls} reviewed in appendix  \ref{app:Canonical_PL}, one obtains that the dual boundary conditions are described by the ``dual'' gluing matrix  $\tilde R$ given by 
\begin{align}\label{eq:tR_gluing}
	\tilde L_-=\tilde R\tilde L_+\,, \quad \tilde{R}=-\tilde{E}^{-1}E_-^{-1}ER(E^T)^{-1}E_0^T\tilde{E}^T\,.
\end{align}
This expression for the dual gluing matrix is precisely the expression obtained in \cite{Albertsson:2006zg}.

%%%%%%%%%%%%%%%%%%%%%%%%%%%%%%%%%%%%%%%%%%%
%%%%%%%%%%%%%%%%%%%%%%%%%%%%%%%%%%%%%%%%%%%
\subsection{Dirac geometry fusion}\label{sec:fusion_via_DS}	
As was already briefly mentioned in section \ref{sec:FM}, it is natural to postulate that the fusion at the level of the target space emulates that of the integral transformation of the defect in the worldsheet picture. The idea that a Fourier-Mukai like operation controls the fusion of target space defects was already put forward for the fusion of topological defects (or bibranes) and boundary conditions in WZW models by \cite{Fuchs:2007fw}. The authors proposed that the submanifold that results from the fusion operation between a D-brane and a topological defect separating two WZW models on a Lie group $\mathcal G$ is given by 
	\begin{align}\label{eq:Fus_manif_FSW}
		\mathcal Y_{B}\star \mathcal Y_D \equiv p_1\left(\mathcal Y_{B}\cap p_2^{-1}(\mathcal Y_{D})\right)\,,
	\end{align}
	where $\mathcal Y_{B}$ is the worldvolume of the bibrane or topological defect in the product space $M_1\times M_2=\mathcal G\times \mathcal G$, $\mathcal Y_D$ is the worldvolume of the D-brane in $M_2=\mathcal G$ and $p_i$ are the usual projections $p_i:M_1\times M_2\rightarrow M_i$. Note that this time the role of the kernel for this Fourier-Mukai inspired operation is played by the worldvolume of the topological defect. In general however, the object resulting from the formula in eq. \eqref{eq:Fus_manif_FSW} is ill-defined: the operation in eq. \eqref{eq:str_fusion_FM} is in general just a subset and not a submanifold. A second drawback is that this proposal for the fusion operation does not include an explicit expression for the two-form $\mathcal F$ living on the fused boundary condition.

In the present article we will take, besides the Fourier-Mukai operation, a second guiding principle to propose an Ansatz for the target space fusion. In previous sections, we stressed that the target space data $(\mathcal Y,\mathcal F)$ capturing both D-branes and defects in some manifold $M$ can be encoded in terms of Dirac structures $L=L(\mathcal Y,\mathcal F)\subset M$. This observation strongly suggest that generalised geometry is the right setting to define the operation of fusion at the level of the target space. Using the Fourier-Mukai transformation as our blue print, we will vindicate in this section a fusion operation of a topological defect with boundary condition using the framework of generalised geometry. 

In section \ref{sec:FM}, we reviewed the relation between the Fourier-Mukai type transformation and the fusion of topological defects with boundary conditions. We would like to translate this transformation as a pull-back of the boundary condition to the product space, its ``convolution'' with the ``kernel'' of the topological defect and subsequence reduction to the second space.  In terms of generalised geometry that means we have to transport Dirac structures from Courant algebroids over the different manifolds. In the present context it will be relevant to restrict to a particular class of Courant algebroids to obtain a well-defined notion of fusion. The following discussion relies on the results of \cite{Severa:2015hta,bursztyn2007reduction,cavalcanti2011generalized} on Courant algebroids and the related reductions.

We will first assume that all three Courant algebroids $E_i$ over $M_i$ for $i=1,2$ and $\mathbbsl E$ over $\mathbbsl M=M_1\times M_2$ are so-called exact Courant algebroids. This class was discovered and classified by \v Severa in \cite{vsevera2017letters,vsevera2001poisson}. These Courant algebroids have the distinguishing property that they are locally isomorphic to the standard Courant algebroid $TM\oplus T^\ast M$, if one allows for the Courant bracket to be twisted by a closed three-form $H$. Exact Courant algebroids are then uniquely determined by a class in $H^3_{\mathrm dR}(M,\mathbb R)$. Making this choice essentially specialises the theories on both sides of the defect to be determined by two-dimensional sigma-models. Indeed one can show that an exact $H$-twisted Courant algebroid $E$ over a manifold $M$ is equivalent to specifying the phase space of a two-dimensional sigma-model where the Kalb-Ramond field has curvature $H$, see e.g. \cite{Severa:2019ddq}. Schematically we now have a Fourier-Mukai-like diagram transporting D-branes and defects as Dirac structure from one Courant algebroid to the other 
\begin{equation}   
\begin{aligned}
\begin{tikzpicture}[x=0.75pt,y=0.75pt,yscale=-1,xscale=1] 
\draw    (120,30.21) -- (93.83,58.71) ;
\draw [shift={(92.48,60.18)}, rotate = 312.55] [color={rgb, 255:red, 0; green, 0; blue, 0 }  ][line width=0.75]    (10.93,-3.29) .. controls (6.95,-1.4) and (3.31,-0.3) .. (0,0) .. controls (3.31,0.3) and (6.95,1.4) .. (10.93,3.29)   ;
%Straight Lines [id:da7238095772608195] 
\draw    (182.15,31.84) -- (201.48,59.18) ;
\draw [shift={(181,30.21)}, rotate = 54.75] [color={rgb, 255:red, 0; green, 0; blue, 0 }  ][line width=0.75]    (10.93,-3.29) .. controls (6.95,-1.4) and (3.31,-0.3) .. (0,0) .. controls (3.31,0.3) and (6.95,1.4) .. (10.93,3.29)   ;

% Text Node
\draw (83,7.08) node [anchor=north west][inner sep=0.75pt]    {$\mathbbsl{E}\cong \mathbb T( M_{1} \times M_{2})_{H_{1} -H_{2}}$};
% Text Node
\draw (181,62.4) node [anchor=north west][inner sep=0.75pt]    {$E_{1} \cong (\mathbb{T} M_{1})_{H_{1}}$};
% Text Node
\draw (48,62.35) node [anchor=north west][inner sep=0.75pt]    {$E_{2} \cong (\mathbb{T} M_{2})_{H_{2}}$};

\end{tikzpicture}\label{eq:diagram_CAs}
\end{aligned}	
\end{equation}
where the right, respectively left, arrows are understood to be the pull-back, resp. the reduction of the  Dirac structure via the associated pull-back/reduction of the Courant algebroids.
To warrant the reduction process to be well-defined, we will in addition demand that these exact Courant algebroids admit the action of a Lie group $\mathcal G$ that is free and transitive such that the associated 1st Pontraygin class vanishes. Such Courant algebroids are known as $\mathcal G$-equivariant (exact) Courant algebroids. In addition, these assumptions guarantee that any Dirac structure in the respective Courant algebroids remain Dirac in the new pull-backed/reduced Courant algebroids  \cite{Severa:2015hta}.

We are left with the task of specifying the ``convolution'' operation, and the relevant ``kernel'', taking place in between the pull-back to the product space and the reduction in eq. \eqref{eq:diagram_CAs}. A simple Ansatz is a natural product on the space of Dirac structures $\mathrm{Dir}(\mathbbsl D)$, which we will denote by $\circledast$ to distinguish it from the fusion operation $\star$.  As we will see that the product $\circledast$ on the space of Dirac structure over $M$, automatically reproduces the desired submanifold resulting from fusion proposed by \cite{Fuchs:2007fw} and given in eq. \eqref{eq:Fus_manif_FSW}. Indeed, take two Dirac structures $L_1\equiv L(\mathcal Y_1,\omega_1),L_2\equiv L(\mathcal Y_2,\omega_2)\subset \mathbb TM$, one can then define a pointwise product\footnote{Dually, we could represent a Dirac structure $L=L(\mathcal Y,\pi)$ where
	$L(\mathcal Y,\pi)=\{X+\pi^\sharp(\eta)+\eta: \eta\in T\mathcal Y^\circ\,,\;X\in T\mathcal Y\}$, where the annihilator space if denoted by a $\circ$.
	For which we can in turn define a product
	$L_1\circledast L_2=\{X_1+X_2+\eta:X_1+\eta\in L_1\,, X_2+\eta\in L_2\}\,.$}
	\begin{align*}
		(L_1\circledast L_2)_\mathbbsl g\equiv L_{1,\mathbbsl g}\circledast L_{2,\mathbbsl g}\subset \mathbb T_p\mathbbsl M\,,\quad\forall \mathbbsl g\in \mathbbsl  M\,,
	\end{align*}
	given explicitly by 
	\begin{align*}
		L(\mathcal Y_1,\omega_1)\circledast L(\mathcal Y_2,\omega_2)&=\{X+\eta+\nu:X+\eta\in L_1, X+\nu\in L_2\}\\
		 &=L(\mathcal Y\subset \mathcal Y_1\cap \mathcal Y_2,\omega_1|_{\mathcal Y}+\omega_2|_{\mathcal Y})\,.
	\end{align*}
If the resulting subbundle is smooth, the product $L_1\circledast L_2$ is also a Dirac structure. See e.g.  \cite{marcut2016introduction} for more details.
	In addition, when $\mathcal Y_1\cap \mathcal Y_2$ is a connected submanifold, we have that $\mathcal Y=\mathcal Y_1\cap \mathcal Y_2$. In general this intersection will have different connected components, possibly of  different dimensions. Note finally that this Dirac structure product $\circledast$ also naturally realises  B-field shifts, since for a closed two-form $B$, we have that $L(\mathbbsl{D}_\mathrm{diag},B)\circledast L(\mathcal Y_1,\mathcal F)= L(\mathcal Y_1,\mathcal F+B)$.

	Summarising, we have the following diagram
	\begin{equation}
\begin{aligned}\label{eq:basic_fusion_diag}
\tikzset{every picture/.style={line width=0.6pt}}         
\begin{tikzpicture}[x=0.75pt,y=0.75pt,yscale=-1,xscale=1]

%Straight Lines [id:da13333440289794485] 
\draw    (120,30.21) -- (93.83,58.71) ;
\draw [shift={(92.48,60.18)}, rotate = 312.55] [color={rgb, 255:red, 0; green, 0; blue, 0 }  ][line width=0.75]    (10.93,-3.29) .. controls (6.95,-1.4) and (3.31,-0.3) .. (0,0) .. controls (3.31,0.3) and (6.95,1.4) .. (10.93,3.29)   ;
%Straight Lines [id:da6720595574283664] 
\draw    (182.15,31.84) -- (201.48,59.18) ;
\draw [shift={(181,30.21)}, rotate = 54.75] [color={rgb, 255:red, 0; green, 0; blue, 0 }  ][line width=0.75]    (10.93,-3.29) .. controls (6.95,-1.4) and (3.31,-0.3) .. (0,0) .. controls (3.31,0.3) and (6.95,1.4) .. (10.93,3.29)   ;

% Text Node
\draw (76,7.4) node [anchor=north west][inner sep=0.75pt]    {$  \bar L(\mathcal Y_B,\mathcal F_B) \circledast  \iota _{\ast }  L(\mathcal Y_1,\mathcal F_1)$};
% Text Node
\draw (52,62.4) node [anchor=north west][inner sep=0.75pt]    {$  L(\mathcal Y_{\mathrm{fus}},\mathcal F_{\mathrm{fus}}) $};
% Text Node
\draw (188,62.4) node [anchor=north west][inner sep=0.75pt]    {$ L(\mathcal Y_1,\mathcal F_1)$};
\end{tikzpicture}
\end{aligned}
\end{equation}
 This proposed fusion operation really contains two pieces of data: on the one hand the resulting submanifold, and on the other the expression for the induced two-form field
\begin{align*}
	\mathcal Y_{\mathrm{fus}}\equiv \mathcal Y_B\star \mathcal Y_1=p(\mathcal Y_B\cap {\iota}(\mathcal Y_1))\subset M_2\,,\quad \mathcal F_{\mathrm{fus}}\equiv (p)_\ast \left(\mathcal F_B  -  {\iota}^\ast(\mathcal F_1)\right)\in \wedge^2 M_2\,,
\end{align*}
where $(\mathcal Y_B,\mathcal F_B)$ is the target space data characterising the topological defect in $M_1\times M_2$ and $(\mathcal Y_1,\mathcal F_1)$ the world-volume data of the D-brane in $M_1$. The notation $\bar L$, denoted the Dirac structure with \textit{minus} the two-form, i.e. $\bar  L(\mathcal Y_B,\mathcal F_B)\equiv  L(\mathcal Y_B,-\mathcal F_B)$. Looking now at the two-form we have that $\mathrm d \mathcal F_B =H_1|_\mathrm{def.}-H_2|_\mathrm{def.}$, where we have restricted the three-forms to the defect location,  for the two-form on the worldvolume of the topological defect. First note that this relation reproduces the fused submanifold given in eq. \eqref{eq:Fus_manif_FSW} of \cite{Fuchs:2007fw}. In addition, by construction we have that $\mathrm d \mathcal F_\mathrm{fus}=H_2$. 

Let us conclude this section with connecting back to the Poisson-Lie T-duality topological defect.
Defining the fusion in this way incidentally matches also exactly the procedure described in \cite{Severa:2018pag} of how Dirac structure transform under T-duality. To see this in more  detail, we have the Poisson-Lie T-duality defects $(\mathcal Y_B,\mathcal F_B)=(\mathbbsl D,\omega_\mathrm{STS})$ and we take a D-brane that is a symplectic leaf associated to the Lie group $\mathcal G$ \cite{klimvcik1996poisson}. To embed a symplectic leaf $\mathcal Y\subset \mathcal G$ in the Drinfel'd double, we can simply multiply every point $g\in \mathcal Y\subset \mathcal G$ with a constant element $\mathbbsl g\in \mathbbsl D$ of the Drinfel'd double. The intersection for this particular case is trivial: $\mathbbsl D \cap \iota \mathcal Y=\iota \mathcal Y\subset \mathbbsl D$. The last step, the projection onto $\tilde{\mathcal G}$, requires one to decompose every point on the embedded surface $\iota \mathcal Y$ 
\begin{align*}
	p_1(\iota \mathcal Y)=\{\tilde h\mid \mathbbsl g g=\tilde h h\;\text{for}\; g\in \mathcal Y\;\text{and}\;h\in \mathcal G, \tilde h\in \tilde{ \mathcal G}\}\,,
\end{align*}
yielding $\mathcal G$-dressing orbits or equivalently the symplectic leaves in the dual Lie group $\tilde{\mathcal G}$ for the Poisson structure $\tilde \Pi$ associated to the double $\mathbbsl D$, in accordance with \cite{klimvcik1996poisson}. In the special case where element $\mathbbsl g=e$ is the identity element, the leaf is dual to a single point. Finally, one can easily check that $\mathcal F_\mathrm{fus}=(\tilde\Pi|_{p_1(\iota \mathcal Y)})^{-1}$ which is well-defined since $p_1(\iota \mathcal Y)$ is precisely a symplectic leaf of the Poisson-Lie structure  $\tilde\Pi$ on $\tilde{\mathcal G}$.

%%%%%%%%%%%%%%%%%%%%%%%%%%%%%%%%%%%%%%%%%%%
%%%%%%%%%%%      Discussion      %%%%%%%%%%	
%%%%%%%%%%%%%%%%%%%%%%%%%%%%%%%%%%%%%%%%%%%	
\section{Discussion}\label{sec:discussion}
In this article, we showed how the most general notion of T-duality, despite its potential disparities with  more conventional notions of T-duality, can in fact be encoded as a defect separating Poisson-Lie T-dual models. Using a target space formulation, we have shown, by direct construction, that Poisson-Lie T-duality can be understood as a space-filling topological defect. This Poisson-Lie T-duality defect is characterised by the natural symplectic structure on the Drinfel'd double associated to the Poisson-Lie T-duality pair, the so-called Semenov-Tian-Shanshy symplectic structure.  In addition, we argued, along the same line as \cite{Sarkissian:2008dq}, that the Fourier-Mukai transform associated to Poisson-Lie T-duality should have the Semenov-Tian-Shansky symplectic structure as kernel. This nicely agrees with the same result obtained by \cite{Arvanitakis:2021lwo} from a very different perspective, that of QP-manifolds. We subsequently demonstrated how the Poisson-Lie T-duality defect consistently yields the correct transformation rules for the boundary condition given by a gluing matrix under generalised T-duality.  

The final part of this paper took a more general look at the target space realisation of fusion, without specialising to the above constructed generalised T-duality defect. Although very well-understood at the level of the worldsheet, the target space realisation of fusion is mired by a veil of mystery and remains largely ill-understood.  A crucial step forwards was made in \cite{Fuchs:2007fw,Kapustin:2009av}, by realising that the fusion process should be similar in spirit to the Fourier-Mukai transform. In the present article, we put forward that a second crucial ingredient is that the worldvolumes of both defects and D-branes can be described within generalised geometry via Dirac structures. Exploiting both insights, we proposed a well-defined operation for the worldvolume fusion using Dirac geometry. The resulting object is always a submanifold carrying a well-defined two-form field whose exterior derivative matches the three-form flux.

Let us conclude with some future directions and open questions. 
It is tempting to believe that, in analogy with topological defects WZW models \cite{Fuchs:2007fw}, the worldvolume of topological defects separating Poisson-Lie symmetry models could be related, not as for WZW models to conjugacy classes, but to dressing orbits of the associated Drinfel'd double. Dressing orbits can be seen as a non-linear generalisation of conjugacy classes associated to a natural action of Poisson-Lie dual pairs. The worldvolume of these bibranes would then be some sort of ``bi-dressing orbits''.

Integrable defects form a special subclass of topological defects. This then begs the question of whether the integrable defects in Yang-Baxter models constructed in \cite{Demulder:2021vus} admit a notion of fusion. The fusion of integrable defects with boundary conditions was e.g. already considered in \cite{Bajnok:2007jg} for Sinh-Gordon and Lee-Yang models. On the other hand, it would be interesting to specialising the duality defects and the associated fusion to particular instances of models which are related by Poisson-Lie T-duality, such as the $\eta$-or $\lambda$-deformations \cite{Klimcik:2002zj,Sfetsos:2013wia,Sfetsos:2015nya} and their boundary conditions \cite{Driezen:2018glg}.

In view of the recent formulation of Poisson-Lie T-duality in terms of QP manifolds, let us comment that Dirac structures on generalised geometry can straightforwardly be lifted to  QP geometry. In the supergeometric setting the Courant algebroid becomes a degree two symplectic N-manifold with Courant bracket and anchor encoded by an Hamiltonian vector field $\Theta$, i.e. $\{\Theta,\Theta\}=0$,  and Dirac structures $L$ of $E$ become Lagrangian submanifolds $\mathcal L$ along which $\Theta|_{\mathcal L}$ is constant \cite{roytenberg1999courant}.

\begin{figure}
	\centering
	\includegraphics[scale=0.21]{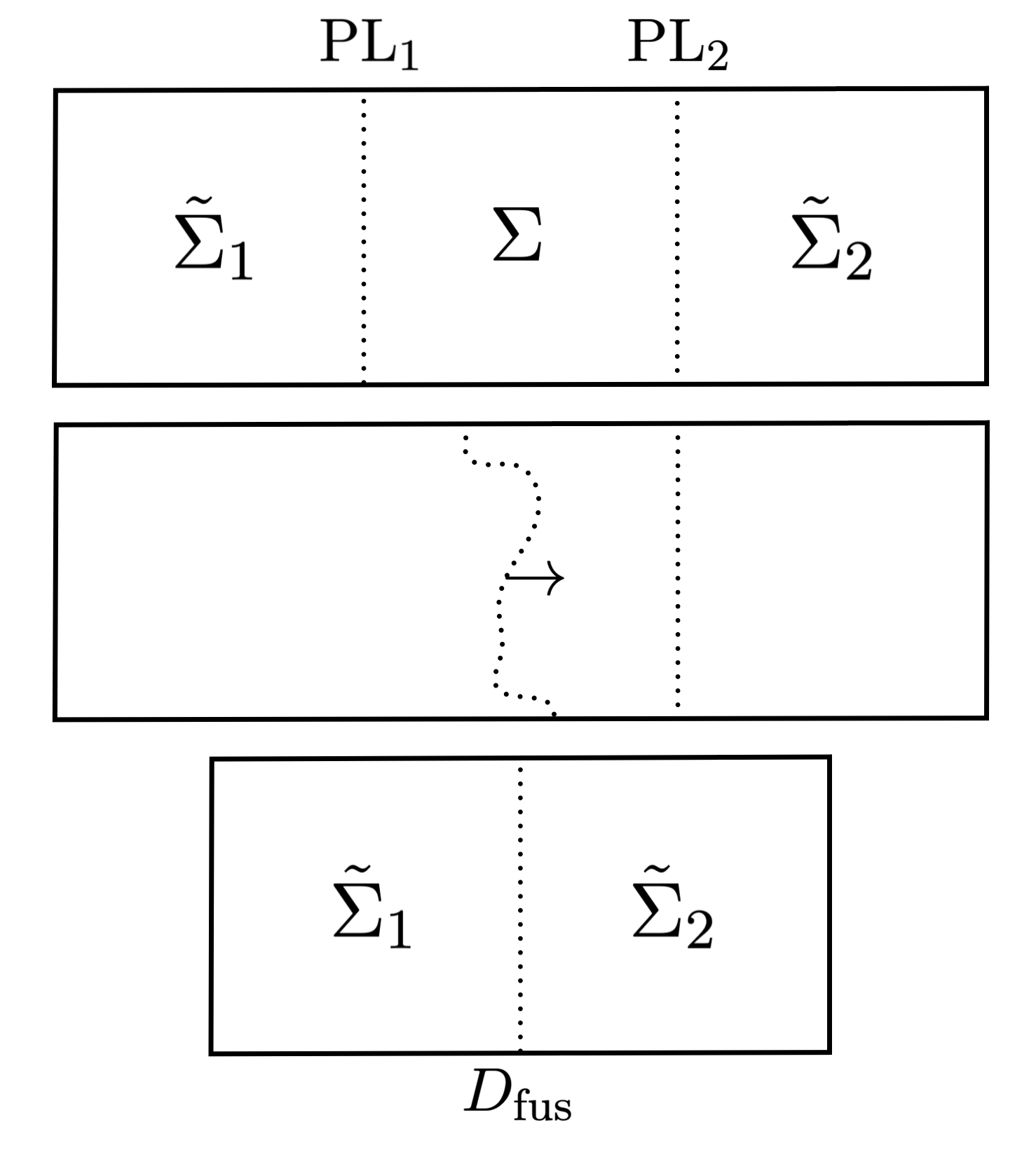}\caption{When an algebra admits two Drinfel'd double one can fuse the associated Poisson-Lie T-duality defects $\mathrm{PL}_i$ separating the sigma-models $\Sigma\rightarrow \mathcal G$ with $\tilde\Sigma_i \rightarrow \tilde{\mathcal G}_i$, for $i=1,2$. }\label{fig:fusion_PLs}
\end{figure}
\noindent

Finally, in the present work, the resulting operation of fusion at the level of the target space was restricted to fusion processes between a topological defect and a boundary condition or D-brane. In fact, the fusion of two topological defects associated to generalised T-dualities appear to be somewhat more subtle and has to be handled with care. To see this, remember that a given Lie group $\mathcal G$ might admit two different Poisson-Lie structures $\mathrm{PL}_{i}$, $i=1,2$, and thus two Poisson-Lie dual groups $\tilde{ \mathcal G}_{i}$ with corresponding doubles $\mathbbsl D_{i}$, see fig. \ref{fig:fusion_PLs}. Assume that we place the worldsheets of Poisson-Lie symmetry on $\mathcal G$ between those with target spaces $\tilde{ \mathcal G}_1$ (on the left) and  $\tilde{\mathcal G}_2$ (on the right), with the worldsheets divided by the respective generalised T-duality defects. Being topological, we can fuse these two topological defects together. More often than not however, the resulting product space $\tilde{\mathcal G}_1\times \tilde{\mathcal G}_2$ will not be a Drinfel'd double and this defect will not encode a (single) Poisson-Lie T-duality. One could however imagine that the fused defect encodes a complicated solution generating transformation that results from applying the two Poisson-Lie T-duality after one another. It has been anticipated in \cite{Bachas:2007td} (see also \cite{Frohlich:2006ch}) that fusion of conformal defects could lead to a solution generating algebra, similar to the role of Ehlers-Geroch transformations in general relativity. This simple reasoning seems to suggest that studying the fusion algebra generated by the Poisson-Lie defect on the target space can potentially enable us to pinpoint crucial differences between Abelian and Poisson-Lie T-duality.

%%%%%%%%%%%%%%%%%%%%%%%%%%%%%%%%%%%%%%%%%%%	
%%%%%%%%%%%%%%%%%%%%%%%%%%%%%%%%%%%%%%%%%%%	
\subsubsection*{Acknowledgments}
We would like to thank Alex Arvanitakis, Chris Blair, Falk Hassler, Dan Thompson, Fridrich Valach, Ilka Brunner, David Osten and Pavol \v Severa for interesting discussions and Dieter L\"ust and Alessandra Gnecchi for support and encouragement. We would also like to thank the organisers and the participants of the ``Integrability, dualities and deformations'' conference at Humboldt Universit\"at zu Berlin for the many interesting discussions.

\newpage
\appendix
%%%%%%%%%%%%%%%%%%%%%%%%%%%%%%%%%%%%%%%%%%%
%%%%%%%%%%%       Appendix       %%%%%%%%%%	
%%%%%%%%%%%%%%%%%%%%%%%%%%%%%%%%%%%%%%%%%%%	
\section{Drinfel'd double conventions and the $\omega_{STS}$}\label{app:STS} 
In this appendix, we set our notation used in the main text as well as sketch the derivation of an alternative expression for the Semenov-Tian-Shansky symplectic form.

A Drinfel'd double $\mathbbsl D$ is an even-dimensional Lie group with Lie algebra $\mathfrak{d}$ (also referred to as the Drinfel'd double) equipped with a non-degenerate symmetric bilinear form $\langle \cdot , \cdot \rangle_\frak{d}$ such that the algebra can be written as $\mathfrak{d} = \mathfrak{g} \oplus \tilde{\mathfrak{g}}$, with $\mathfrak{g}$ and $\tilde{\mathfrak{g}}$ maximally isotropic subalgebras with respect to $\langle \cdot , \cdot \rangle_\frak{d}$. The generators of the Drinfel'd double $\frak d$ are denoted by $\mathbbsl{T}_A=(T_a,\tilde{T}^a)$ with $T_a,\tilde{T}^a$ the generators of the maximally isotropic subalgebras $\mathfrak{g}, \tilde{\mathfrak{g}}$. The triple $(\mathfrak{d},\mathfrak{g},\tilde{\mathfrak{g}} )$ corresponding to this three quantities is called a Manin triple. The isotropy condition means that 
\begin{align*}
\langle T_a,T_b \rangle_\frak{d} = 0\,, \quad \langle \tilde{T}^a,\tilde{T}^b \rangle_\frak{d} =0\,, \quad \langle T_a , \tilde{T}^b \rangle_\frak{d} = \delta_a^b\,,
\end{align*}
and the maximality conditions requires the Lie algebra to be half of the  dimension of the corresponding Drinfel'd double, i.e. $\mathrm{dim}\, \frak d= 2\,\mathrm{dim}\, \frak g= 2\,\mathrm{dim}\, \tilde{\frak{g}}$.
The Lie groups $\mathcal G$ and $\tilde{\mathcal G}$ obtained after exponentiating  the Lie algebras $\frak g$ and $\tilde{\frak{g}}$ carry a natural Poisson-Lie structure $\Pi$ and $\tilde \Pi$ compatible with the group multiplication. 
For group elements $g \in \mathcal G= \exp \mathfrak{g}, \tilde{g} \in \tilde{\mathcal G} = \exp \tilde{\mathfrak{g}}$, we define the adjoint action matrices by
\begin{align*}
g^{-1} T_a g &= a_a^{\ b}[g] T_b\,, \quad g^{-1}\tilde{T}^b g = b^{ab}[g] T_b + (a[g]^{-1})_b^{ \ a} \tilde{T}^b\,, \\
\tilde{g}^{-1} \tilde{T}^a \tilde{g} &= \tilde{a}^a_{\ b}[\tilde{g}] \tilde{T}^b\,, \quad \tilde{g}^{-1} T_a \tilde{g} = \tilde{b}_{ab}[\tilde{g}] \tilde{T}^b + (\tilde{a}[\tilde{g}]^{-1})^b_{ \ a} T_b\,.
\end{align*}
In terms of these adjoint action matrices, the Poisson-Lie structures can easily be computed using
\begin{align}
\Pi^{ab}[g] = b^{ca}[g]a_c^{\ b}[g]\,, \quad \tilde{\Pi}_{ab}[\tilde{g}] = \tilde{b}_{ca}[\tilde{g}]\tilde{a}^c_{\ b}[\tilde{g}]\,.
\end{align}
One can show by direct computation that, as should be for Poisson structures, these bivectors are antisymmetric $\Pi^T[g] = - \Pi[g]$ and idem for the tilde object.

We show now how one can get the particular form of the Semenov-Tian-Shansky symplectic form $\omega_\mathrm{STS}$ we use in order to define the defect from its standard expression given in \cite{semenov1985dressing,Alekseev:1993qs}
\begin{align}\label{eq:app_def_STS}
2 \omega_\mathrm{STS} = r^a(g) \wedge \tilde{l}_a(\tilde{h}) - \tilde{r}_a(\tilde{g}) \wedge l^a(h)\,.	
\end{align}
The expression for the Semenov-Tian-Shansky we need in the main text closely resembles an expression first derived in \cite{Arvanitakis:2021lwo}. 
In order to do so we first need to collect a number of identities relating the different decompositions of the group element on the Drinfel'd double $\mathbbsl D=\mathcal G\tilde{\mathcal G}$. Recall that an element $\mathbbsl{g}$ on the double $\mathbbsl D$ can be written in two different ways
\begin{align*}
\mathbbsl{g} = \tilde{h} g \quad \text{or} \quad \mathbbsl{g} = h \tilde{g}\,,\quad \text{with}\;\,\,g,h\in \mathcal G\,,\ \tilde g,\tilde h\in \tilde{\mathcal G}\,.
\end{align*}
First we compare the adjoint action on the double in these two different parameterisations
\begin{align*}
\mathbbsl{a}[\mathbb{g}=\tilde{h}g] T_a &= (\tilde{h}g)^{-1}T_a \tilde{h}g=g^{-1}\left(\tilde{b}_{ab}[\tilde{h}] \tilde{T}^b+ (\tilde{a}^{-1})^b_{\ a}[\tilde{h}]T_b \right)g\\
&=\tilde{b}_{ab}[\tilde{h}]\left( b^{bc}[g]T_c +(a^{-1})_c^{\ b} [g] \tilde{T}^c\right) + (\tilde{a}^{-1})^b_{\ a}[g] a_b^{\ c}[g]T_c\,,
\end{align*}
and for the alternative decomposition of the Drinfel'd double element
\begin{align*}
\mathbbsl{a}[\mathbb{g}=h\tilde{g}] T_a &=a_a^{\ b}[h]\left( \tilde{b}_{bc}[\tilde{g}] \tilde{T}^c + (\tilde{a}^{-1})^c_{\ }[\tilde{g}] T_c \right)\,.
\end{align*}
Collecting the terms in valued in $\frak g$,  we get the identifications
\begin{align}
a_a^{\ b}[h](\tilde{a}^{-1})^c_{\ b}[\tilde{g}] &= \tilde{b}_{ab}[\tilde{h}] b^{bc}[g] + (\tilde{a}^{-1})^b_{\ a} [\tilde{h}] a_b^{\ c}[g] \label{eq:APP_id1}\\
\tilde{b}_{ab}[\tilde{h}](a^{-1})_c^{\ b}[g] &= a_a^{ \ b }[h] \tilde{b}_{bc}[\tilde{g}]\label{eq:APP_id2}\,.
\end{align}
Doing the same for the terms in valued in the dual algebra $\frak g$, leads to
\begin{align}
\tilde{a}^a_{\ b}[\tilde{h}] (a^{-1})^b_{\ c}[g] &= b^{ab}[h] \tilde{b}_{bc}[\tilde{g}] + (a^{-1})_b^{\ a}[h] \tilde{a}_c^{\ b}[\tilde{g}]\label{eq:APP id3}\\
\tilde{a}^a_{\ b}[\tilde{h}]b^{bc}[g] &= b^{ab}[h](\tilde{a}^{-1})^c_{\ b}[\tilde{g}]\label{eq:APP id4}\,.
\end{align}
We proceed analogously with the invariant forms $\mathbbsl{L}= \mathbbsl{g}^{-1} \mathrm d \mathbbsl{g}$ and $\mathbbsl{R} =  \mathrm d \mathbbsl{g} \mathbbsl{g}^{-1}$.
\begin{align*}
\mathbbsl{L}[\mathbb{g}=\tilde{h}g] &= g^{-1} \tilde{l}(\tilde{h}) g + l(g) = \tilde{l}_a(\tilde{h}) ( b^{ab}[g]T_b + (a^{-1})_b^{\ a}[g]\tilde{T}^b) + l^b(g) T_b\\
\mathbbsl{L}[\mathbb{g}=h \tilde{g}] &= \tilde{g}^{-1} l(h) \tilde{g} + \tilde{l}(\tilde{g}) = l^a(h)(\tilde{b}_{ab}[\tilde{g}]\tilde{T}^b + (a^{-1})^b_{\ a}[\tilde{g}]T_b + \tilde{l}_b(\tilde{g}) \tilde{T}^b
\end{align*}
Therefore we have
\begin{align*}
\tilde{l}_a(\tilde{h}) b^{ab}[g]+l^a(g) & =l^a(h) (a^{-1})[\tilde{g}]\,\quad 
\tilde{l}_a(\tilde{h})(a^{-1})_b^{\ a}[g] =  l^a(h)\tilde{b}_{ab}[\tilde{g}] + \tilde{l}_a(\tilde{g})%\label{eq:APP_id5_and_id6}
\end{align*}
as well as from $\mathbb{R}$
\begin{align*}
r^b(h) &= r^a(g)\tilde{a}^b_{\ a}[\tilde{h}]- \tilde{r}_a(\tilde{g}) b^{ab}[h^{-1}]\,,\quad 
\tilde{r}_b(\tilde{h})=\tilde{r}_a  (\tilde{g})a_b^{\ a}[h] - r^a(g)\tilde{b}_{ab}[\tilde{h}^{-1}]   \,,
\end{align*}
which implies the following relation between left- and right-invariant Maurer-Cartan fields 
\begin{align}
	  l^a(h)&= r^d(g)\tilde{a}^b_{\ d}[\tilde{h}] a_b^{\ a}[h] - \tilde{r}_d(\tilde{g}) b^{db}[h^{-1}] a_b^{\ a}[h]\,,\label{eq:APP_id7} \\ 
	  \tilde{l}_a(\tilde{h}) &= \tilde{r}_d  (\tilde{g})a_b^{\ d}[h] \tilde{a}^b_{\ a}[\tilde{h}] - r^d(g)\tilde{b}_{db}[\tilde{h}^{-1}] \tilde{a}^b_{\ a}[\tilde{h}]\,.\label{eq:APP_id8}
\end{align}
Lastly we also need to find an expression for the matrix $C$ featured in the Poisson-Lie T-duality transformation rules. In order to do so, we start from equation \eqref{eq:APP_id1} and apply eq. \eqref{eq:APP_id2} as well as $\Pi^{ab}[g] = b^{ca}[g]a_c^{\ b}[g], \tilde{\Pi}_{ab}[\tilde{g}] = \tilde{b}_{ca}[\tilde{g}] \tilde{a}^c_{\ b}[\tilde{g}]$ to get
\begin{align*}
a_a^{\ b}[h](\tilde{a}^{-1})^c_{\ b}[\tilde{g}] &= \tilde{b}_{ab}[\tilde{h}] b^{bc}[g] + (\tilde{a}^{-1})^b_{\ a} [\tilde{h}] a_b^{\ c}[g]\,.
\end{align*}
This relation then allows one to write the following expression for the matrix $C$ in terms of the adjoint maps
\begin{align}
 C_c^{\ h} \equiv  (\delta^c_h-\tilde{\Pi}_{hl}[\tilde{g}] \Pi^{lc}[g])^{-1} &= (a^{-1})_c^{\ b}[g] (\tilde{a})^a_{\ b}[\tilde{h}] a_a^{\ e}[h](\tilde{a}^{-1})^h_{\ e}[\tilde{g}] \label{eq:APP_id9}\,.
\end{align}
Having collected all the necessary expressions at hand, we can finally compute the relevant expression for the Semenov-Tian-Shansky symplectic form. Starting from the defining expression for the symplectic form in equation \eqref{eq:app_def_STS} and recalling that $\tilde{r}_a=\tilde{l}_b (\tilde{a}^{-1})^b_{\ a}[\tilde{g}] , r^a= l^b (a^{-1})_b^{\ a}$,  we can consecutively rewrite the symplectic structure as
\begin{align*}
2 \omega_{STS} &= l^b(g) (a^{-1})_b^{\ a}[g] \wedge\left(\tilde{r}_d  (\tilde{g})a_c^{\ d}[h] \tilde{a}^c_{\ a}[\tilde{h}] - r^d(g)\tilde{b}^{dc}[\tilde{h}^{-1}] \tilde{a}^c_{\ a}[\tilde{h}] \right)\\
&\quad - \tilde{l}_b(\tilde{g}) (\tilde{a}^{-1})^b_{\ a} [\tilde{g}] \wedge  \left( r^d(g)\tilde{a}^c_{\ d}[\tilde{h}] a_c^{\ a}[h] - \tilde{r}_d(\tilde{g}) b^{dc}[h^{-1}] a_c^{\ a}[h]\right) \\
			&= l^b(g)  \wedge \left((a^{-1})_b^{\ a}[g] \tilde{a}^c_{\ a}[\tilde{h}] a_c^{\ d}[h] (\tilde{a}^{-1})^e_{\ d}[\tilde{g}] \right)\tilde{l}_e(\tilde{g})\\
	& \quad - l^b(g)  \wedge \left((a^{-1})_b^{\ a}[g]\tilde{a}^c_{\ a}[\tilde{h}] a_c^{\ d}[h] \tilde{b}_{de}[\tilde{g}] \right) l^e(g)\\
	& \quad - \tilde{l}_b(\tilde{g})  \wedge \left( (a^{-1})_e^{\ d}[g] \tilde{a}^c_{\ d}[\tilde{h}] a_c^{\ a}[h] (\tilde{a}^{-1})^b_{\ a}[\tilde{g}] \right) l^e(g) \\
	&\quad + \tilde{l}_b(\tilde{g})  \wedge \left(\tilde{a}^c_{\ d}[\tilde{h}]b^{de}[g] a_c^{\ a}[h](a^{-1})^b_{\ a}[\tilde{g}] \right)\tilde{l}_e(\tilde{g})\,,
\end{align*}
Here we used equations (\ref{eq:APP_id7}) and (\ref{eq:APP_id8}) to obtain the first line. The second equality results from using (\ref{eq:APP_id2}) and (\ref{eq:APP id4}). This last line, then finally leads, looking back at eq.  (\ref{eq:APP_id9}), to the expression relevant for the symplectic form on the double used in section \ref{sec:top_def_product_space}:
\begin{align}\label{eq:STS_worked_out_app}
	2 \omega_{STS} = 2 l^b(g)  \wedge C_b^{\ e}\tilde{l}_e(\tilde{g}) +   l^b(g)  \wedge C_b^{\ m} \tilde{\Pi}_{me}[\tilde{g}] l^e(g)-\tilde{l}_b(\tilde{g})  \wedge \tilde{C}^b_{\ m} \Pi^{me}[g]  \tilde{l}_e(\tilde{g}) \,. 
\end{align}
Finally, note that here we have used a different decompositions of elements in the double as compared to \cite{Arvanitakis:2021lwo}, where $\mathbbsl{g} = \tilde{g} g $ or $\mathbbsl{g}=g^\prime \tilde{g}^\prime$. Substituting $\tilde{g} \to \tilde{h}, g^\prime \to h , \tilde{g}^\prime \to \tilde{g}$ and using some of the above identities one can recover their parametrization, which they propose as a kernel for the Fourier-Mukai transform.

%%%%%%%%%%%%%%%%%%%%%%%%%%%%%%%%%%%%%%%%%%%	
%%%%%%%%%%%%%%%%%%%%%%%%%%%%%%%%%%%%%%%%%%%	
%%%%%%%%%%%%%%%%%%%%%%%%%%%%%%%%%%%%%%%%%%%	
\section{Canonical transformations for Poisson-Lie T-duality}\label{app:Canonical_PL}
Poisson-Lie T-duality was identified as being a canonical transformations in \cite{Klimcik:1995ux,Klimcik:1995dy,Sfetsos:1996xj,Sfetsos:1997pi} and when relating the left-invariant Maurer-Cartan form and its dual counterpart it takes the form            
\begin{align}\label{eq:canonical_transfo_Ls}
(E_0^\mp)^{-1} E^{\mp}L_\pm =  \pm\tilde{E}^\mp \tilde{L}_\pm
\end{align}
with $L_+= L^a_+ T_a = l^a_i \partial X^i T_a$ and $L_-= L^a_- T_a = l^a_i \bar{\partial} X^i T_a$ (mutatis mutandis for the tilde variables $\tilde{L}_\pm$), and the background fields are given by the usual
\begin{align}
E^\pm&=((E_0^\pm)^{-1}\pm \Pi)^{-1}\,,\quad \tilde{E}^\pm = (E_0^\pm \pm \tilde{\Pi})^{-1}\,.
\end{align}
Note that in this notation we have $(E_0^+)^T=E_0^-$.

Here we rewrite the canonical transformation for Poisson-Lie T-duality into a form amenable to show the topological nature of the Poisson-Lie defect in section \ref{sec:Lagr_def_no_spect}. The canonical relations can be written as 
\begin{gather}\label{eq:canonical_PL_app}
\begin{aligned}
		C^{-1}(E+C\tilde{\Pi})\tilde{C}^{-1}(\tilde{E}+\tilde{C}\Pi)&=1\,,\\
	C^{-1}(E^T - C\tilde{\Pi})\tilde{C}^{-1}(\tilde{E}^T-\tilde{C}\Pi)&=1\,.
\end{aligned}
\end{gather}
From these relations we want to extract the dual background fields $\tilde G, \tilde B$ in terms of the original fields $ G,  B$. Take for example the first relation in eq. \eqref{eq:canonical_PL} which is equivalent to 
\begin{align*}
	 \left( \tilde{C}^{-1} \tilde{G} C^{-1} + \tilde{C}^{-1}(\tilde{B}+\tilde{C} \Pi) C^{-1} \right)^{-1} &= G + (B+C \tilde{\Pi})\,.
\end{align*}
Using that the first, respectively second, term (inside the brackets on the left-hand side) on each side is symmetric, respectively antisymmetric, one obtains the relations\footnote{For $S,\tilde S$ symmetric invertible matrices and $A,\tilde A$ antisymmetric matrices $(\tilde S+\tilde A)^{-1}=(S+A)$ then 
\begin{align*}
	\tilde S^{-1}\tilde A=-AS^{-1}\,,\quad \tilde S=(S+AS^{-1}A)^{-1}\,,
\end{align*}
when $A$ is also invertible one has also that 
$\tilde A=(A+SA^{-1}S)^{-1}$. We will however not require the B-field $B$ to be invertible, since only the first two relations will be needed.}
\begin{gather}\label{eq:dual_bckgrd_ito_original}
	\begin{aligned}
		\tilde{C}^{-1}\tilde{G}C^{-1}&=\left(G-(B+C \tilde{\Pi})G^{-1}(B+C \tilde{\Pi})\right)^{-1}\\
C\tilde{G}^{-1}(B+C \tilde{\Pi}) C^{-1} &= -(B+C\tilde{\Pi}) G^{-1}\,.
	\end{aligned}
\end{gather}
When including spectator variables $Y^i$, the canonical transformations read \cite{Sfetsos:1997pi}
\begin{equation}
(E_0^\mp)^{-1} E^\mp (L_\pm \pm \Pi F^\mp \partial_\pm Y) = \pm \tilde{E}^\mp (\tilde{L}_\pm \mp F^\mp \partial_\pm Y)\,.
\end{equation}
where again
\begin{align}
E^\pm&=((E_0^\pm)^{-1}\pm \Pi)^{-1}\,,\quad \tilde{E}^\pm = (E_0^\pm \pm \tilde{\Pi})^{-1}\,.
\end{align}
and the background fields with spectator directions are restricted to be of the form \cite{Sfetsos:1997pi} (setting $E \equiv E^+$ for the sake of clarity)
\begin{align}
\begin{matrix*}[l]
E_{a\mu} = E_{ab} ((E_0)^{-1})^{bc} F_{c\mu}\,,  & E_{\mu \nu} = F_{\mu \nu} - F_{\mu a}\Pi^{ab}E_{bc}((E_0)^{-1})^{cd}F_{d \nu}\,, \\
\tilde{E}^{a}_{\ \mu} = \tilde{E}^{ab}F_{b \mu}\,, & \tilde{E}_{\mu \nu} = F_{\mu \nu} - F_{\mu a}\tilde{E}^{ab}F_{b \nu}\,,
\end{matrix*}
\end{align}
with $F_{\mu \nu}, F_{a \mu}$ arbitrary matrices depending on the choice of  background configuration chosen.
The canonical transformation for fields not involving any spectator direction are still given by  (\ref{eq:canonical_PL}) while for mixed coordinates we have \cite{Sfetsos:1997pi}
\begin{gather}
\begin{aligned}
		\tilde{E}^d_{\ \mu} &= (\tilde{C}(E+C\tilde{\Pi})^{-1})^{d c}E_{c\mu}\,,\quad 
\tilde{E}^{\ d}_\mu = (\tilde{C}(E^T-C\tilde{\Pi})^{-1})^{d c}E_{c\mu}\,.
	\end{aligned}
\end{gather}

%%%%%%%%%%%%%%%%%%%%%%%%%%%%%%%%%%%%%%%%%%%	
%%%%%%%%%%%%%%%%%%%%%%%%%%%%%%%%%%%%%%%%%%%	
%%%%%%%%%%%%%%%%%%%%%%%%%%%%%%%%%%%%%%%%%%%	
\section{Reminder of generalised geometry }\label{app:GG}
In this appendix, we collect a number of relevant definitions and results in generalised geometry and Courant algebroid theory. See e.g. \cite{gualtieri2011generalized,Severa:2015hta} for a pedagogical or more complete exposition.

%%%%%%%%%%%%%%%%%%%%%%%%%%%%%%%%%%%%%%%%%%%	
\paragraph{Isotropic and Lagrangian subspaces.} For $V$ a real vector space equipped with a real bilinear form $\langle\cdot,\cdot\rangle$, a subspace $L$ of $V\otimes V^\ast$, where $V^\ast$ is linear vector space dual to $V$, is called isotropic if $\langle X, Y\rangle=0$ for all $X,Y\in L$. It will become clear in the following that so-called maximal isotropic subspace play a special role. Since the signature of the bilinear form is $(\dim V,\dim V)$, a maximal isotropic subspace (or Lagrangian subspace) of $V$ is of dimension $\dim V$. Lagrangian subspaces $L\subset V\otimes V^\ast$ for a linear vector space $V$ are equivalent to picking a subspace $\Delta \subset V$ together with a two-form $F\in \wedge^E \Delta^\ast$ via the identification
\begin{align*}
	L=L(\Delta,\epsilon)=\{X+\eta\in \Delta\oplus V^\ast\;:\; \iota^\ast \eta=\imath_XF\}\,,
\end{align*}
where $\iota:\Delta\hookrightarrow V$ is the inclusion map. An important property of Lagrangian subspace of linear vector spaces is that $B$-field transformations of such Lagrangian subspace $L$, i.e.
\begin{align*}
	e^BL(\Delta,F)=L(\Delta,F+\iota^\ast B)\,,\quad B\in \wedge^2 V\,,
\end{align*}
does not change the projection $\Delta$ onto the vector space $V$. Note that using $B$-field transformation any Lagrangian subspace $L$ can be brought to the form $L(\Delta,0)$. Acting with a $\beta$-transformation will in general modify $\Delta$ and its dimension.

%%%%%%%%%%%%%%%%%%%%%%%%%%%%%%%%%%%%%%%%%%%	
\paragraph{Courant algebroids.} A Courant algebroid is a bundle $E$ over an $m$-dimensional smooth manifold $M$ equipped with a bilinear form $\langle \cdot,\cdot \rangle$ of signature $(m,m)$. In addition it comes with an anchor map $a:E\rightarrow TM$ , that allows to map any section on $E$ to the more familiar notion of vector fields on the tangent bundle $TM$ of $M$ and a bracket called the Courant bracket $\llbracket\cdot,\cdot\rrbracket_C$. Contrary to a Lie bracket, this bracket need not be skew-symmetric. Instead one requires the following axioms
\begin{align*}
	\llbracket u,\llbracket v,w\rrbracket \rrbracket &= \llbracket \llbracket u,v\rrbracket , w\rrbracket + \llbracket v,\llbracket u, w\rrbracket \rrbracket \,,\\
	[u,fv]&=f[u,v]+(\rho(u)f)v\,,\\
	\rho(u)(v,w)&= \langle [u,v],w\rangle +\langle v,[u,w]\rangle\,,\\
	[u,v]+[v,u]&=\rho^T\mathrm d \langle u,v\rangle\,,
\end{align*}
for section $u,v,w$ of $E$, $f\in C^\infty(M)$ a continuous function and $\rho^T:T^\ast M\rightarrow E$ is the transpose of the anchor map, using the  non-degenerate innerproduct $\langle\cdot,\cdot\rangle$ to identify $E$ with $E^\star$.  The quadruple $(E,\langle \cdot,\cdot\rangle, a, \llbracket\cdot,\cdot\rrbracket_C)$ is called a Courant algebroid over $M$.
 We will now consider a smooth manifold $M$ together with its generalised tangent bundle $TM\oplus T^\ast M$.

%%%%%%%%%%%%%%%%%%%%%%%%%%%%%%%%%%%%%%%%%%%	
\paragraph{Exact Courant algebroids.}  Take $E$ a Courant algebroid over a manifold $M$ with anchor map $a$. One then has automatically that $a\circ a^T=0$ leading to the chain complex
\begin{align*}
	0\rightarrow T^\ast M \xrightarrow[]{a^T} E\xrightarrow[]{a}TM\rightarrow 0\,.
\end{align*}
When this sequence is exact, then $E$ is called an exact Courant algebroid. The reason why this forms a distinguished subclass amongst Courant  algebroids is that exact Courant algebroids are classified by closed three-forms $H\in \Omega^3(M)$, which in the sigma-model setting coincides with the three-form flux $\mathrm d B=H$ of the Kalb-Ramond two-form $B$. To extract such three-from for a given exact Courant algebroid one uses that since the above sequence is exact, for any Lagrangian subbundle $L$ of $E$ the map $a|_L:L\rightarrow TM$ is an isomorphism. This allows us to make a choice for the embedding $\sigma :TM\rightarrow E$ in the Courant algebroid by demanding that its image is the Lagrangian subbundle $L$. Then 
\begin{align*}
	H(X,Y,Z)\equiv \langle [\sigma(X),\sigma(Y)],\sigma(Z)\rangle\,,
\end{align*}
defines a closed three-form on $M$. Exact Courant algebroids can  be identified with the canonical generalised tangent bundle $TM\oplus T^\ast M$ via the map $\sigma\oplus a^T$ at the cost of twisting the Courant bracket by the three form field $H$:
\begin{align*}
	\llbracket (X,\xi),(Y,\eta)\rrbracket_{H}=([X,Y],L_X\eta-\imath_Y\mathrm d \xi +H(X,Y,\cdot))\,.
\end{align*}
Conversely, for any closed three-form $H\in \Omega^3(M)$, the above bracket turns the canonical generalised tangent bundle $TM\oplus T^\ast M$ into an exact Courant algebroid. Since changing the choice of splitting map $\sigma$ by a two-form $\epsilon\in\Omega^2(M)$, the three-form $H$ is modified by the shift $H+\mathrm d\epsilon$, we have the exact Courant algebroids over a manifold $M$ are uniquely determined by $H^3(M,\mathbb R)$.

\paragraph{ $\mathcal G$-equivariant Courant algebroids.} Take $\frak g$ a Lie algebra equipped with an invariant bilinear symmetric form and denote by $\mathcal G$ the corresponding Lie group. A $\frak g$-equivariant Courant algebroid over a manifold $M$ is a Courant algebroid $E$ together with an injective linear map $\chi:\frak g\rightarrow\Gamma (E)$ which verifies
\begin{align*}
	[\chi(X),\chi(Y)]=[X,Y]\,,\qquad \langle \chi(X),\chi(Y)\rangle=\langle X,Y\rangle\,, 
\end{align*}
for all $X,Y$ in $\frak g$. This definition guarantees that the Courant algebroid $E$ admits the action of the algebra $\frak g$, in turn inducing an action of $\frak g$ on the manifold $M$. When the action of the Courant algebroid $E$ integrates to an action of the associated connected Lie group $\mathcal G$, the algebroid $E$ is called $\mathcal G$-equivariant.  

One can show that when a Lie group $\mathcal G$ acts freely and transitively over $M$, there exists a $\mathcal G$-equivariant exact Courant algebroid over $M$ if and only if the first Pontryagin class $[\langle F,F\rangle_\frak{g}]$ of the bundle $M\rightarrow M/ \mathcal G$ vanishes. 

%%%%%%%%%%%%%%%%%%%%%%%%%%%%%%%%%%%%%%%%%%%	
\paragraph{Dirac structures. }
Dirac structures generalise (or unify) the notion of Poisson structures, symplectic structures and foliations using the framework of generalised geometry. The crucial observation is that these three examples can be equivalently described by a Lagrangian subbundle that is closed with respect to the Courant bracket. 
Indeed, to each symplectic structure $\omega$ or Poisson structure $\pi$ define the associate bundle maps  
\begin{align}
	\omega^\sharp&:TM\rightarrow T^\ast M:\; X\mapsto \imath_X\omega\, \quad \text{or}\quad 
	\Pi^\sharp:T^\ast M\rightarrow T M:\; \alpha\mapsto \imath_\alpha\Pi\,,
\end{align}
then their respective graphs 
\begin{align*}
	L_\Pi&=  \text{graph}(\pi^\sharp)=\{(\pi^\sharp(\xi),\xi)\mid \xi\in \mathbb T^\ast M\}\,,\\
	L_\omega&=\text{graph}(\omega^\sharp)=\{(X,\omega^\sharp(X))\mid X\in \mathbb TM\}\,,
\end{align*}
generate Lagrangian subbundles of $\mathbb TM=TM\oplus T^\ast M$. These subbundles are in addition closed under the Courant bracket by virtue of $\Pi$ satisfying the Schouten bracket, respectively $\omega$ being closed.
 
More generally, a Dirac structure $L$ on $M$ is a subbundle $L\subset \mathbb TM$ such that
\begin{itemize}
\item[(i)] $L=L^\perp$ with respect to $\langle\cdot,\cdot\rangle$\,,
\item[ii)] $L$ is involutive with respect to the Courant bracket: $\llbracket \Gamma(L),\Gamma(L)\rrbracket \subset \Gamma(L)$.
\end{itemize}

%%%%%%%%%%%%%%%%%%%%%%%%%%%%%%%%%%%%%%%%%%%
%%%%%%%%%%%%    Bibliography     %%%%%%%%%%	
%%%%%%%%%%%%%%%%%%%%%%%%%%%%%%%%%%%%%%%%%%%
\bibliographystyle{JHEP}
\bibliography{PL_top_defectsv2.bib}

\end{document}